\documentclass[journal]{IEEEtran}
\usepackage{setspace}
\usepackage{graphicx}
\graphicspath{{./Figs/}}
\usepackage{epstopdf}
\usepackage[cmex10]{amsmath}
\usepackage{amssymb}
\usepackage{mathtools}
\usepackage{array}
\usepackage{url}
\usepackage{cite}

\newtheorem{definition}{\bf Definition}

\begin{document}
\title{Incentivizing Spectrum Sharing via Subsidy Regulations}

\author{Arvind~Merwaday,
		Murat~Yuksel,
		Thomas~Quint,
        Ismail~G\"uven\c{c},
        Walid~Saad,
        and~Naim~Kapucu
\thanks{A. Merwaday and I. Guvenc are with the Department of Electrical and Computer Engineering, Florida International University, Miami, FL, 33172 USA e-mail: (see https://sites.google.com/site/iguvenc/home).}
\thanks{M. Yuksel is with the Department of Computer Science and Engineering, University of Nevada, Reno, Reno, NV, 89557 USA.}
\thanks{T. Quint is with the Department of Mathematics and Statistics, University of Nevada, Reno, Reno, NV, 89557 USA.}
\thanks{W. Saad is with Wireless@VT, Bradley Department of Electrical and Computer Engineering, Virginia Tech, Blacksburg, VA, 24061 USA.}
\thanks{N. Kapucu is with the School of Public Administration, University of Central Florida, Orlando, FL, 32816 USA.}
}


\maketitle

\begin{abstract}
Traditional regulatory methods for spectrum licensing have been recently identified as one of the causes for the under-utilization of the valuable radio spectrum. Governmental agencies such as the Federal Communications Commission (FCC) are seeking ways to remove stringent regulatory barriers and facilitate broader access to the spectrum resources. The goal is to allow for an improved and ubiquitous sharing of the precious radio spectrum between commercial service providers. 

In this paper, we propose a novel noncooperative game theoretic approach, to show how to foster more sharing of the radio spectrum via the use of regulatory power. We define a two stage game in which the government regulators move first, followed by the providers. The providers are incentivized by lower spectrum allocation fees from the regulators in return for \emph{proof-of-sharing}. The providers are offered discounted spectrum bands, potentially at different locations, but will be asked to provide coverage to users that are not subscribed to them so as to maintain their subsidy incentives from the government. In a simplification of the model, analytical expressions for the providers' perfect equilibrium strategies are derived, and we argue for the existence of the government's part of a perfect equilibrium. Our analysis shows that through subsidization, the government can provide small service providers a fair chance to compete with the large providers, thereby avoiding monopolization in the market. 
\end{abstract}


\section{Introduction}
\IEEEPARstart{W}{ith} the continuous increase in mobile traffic, users and mobile communication services are facing the problem of spectrum shortages. An 11-fold increase in global mobile data traffic is expected between 2013 and 2018 \cite{2014-cisco-white-paper}. Governments around the world are trying to find ways in which more spectrum can be made available not only for mobile use, but also for other services that involve the use of wireless broadband technologies such as weather forecast and surveillance~\cite{power-sharing-2013}. Moreover, wireless networking technologies are becoming a more critical platform for disaster management and public safety applications, which mandates better communications and interoperability by effectively exploiting under-utilized spectrum resources following a disaster~\cite{2012-mousa-challenges}. Conceptualizing the next generation cellular technology shows that improvement in the spectrum regulations will become a critical consideration to meet the future traffic demands \cite{WhatWill5GBe}.

Developing an efficient spectrum policy and new regulations regarding spectrum use are essential in this day and age. Every three years, representatives and delegates from many nations get together to discuss the future of spectrum policies and ways to make spectrum use more efficient and equitable. The World Radio communication Conference (WRC), supported by the United Nations, is usually held in the International Telecommunication Union (ITU) head-quartered in Geneva~\cite{2012-clegg-results}.

The main issue and challenge ahead is to create additional spectrum capacity. Regulators have three policy directions to increase the spectrum capacity: (i) increasing availability and access to radio spectrum for wireless broadband via allocating additional spectrum; (ii) reassigning spectrum to new users; and (iii) opening up spectrum for unlicensed use. Other policy options that may be employed to increase spectrum capacity include: (i) sharing the wireless network infrastructure; (ii) changing the cost structure of spectrum access; (iii) moving to more spectrum-efficient technologies; and (iv) sharing spectrum. Most of these approaches involve sharing the spectrum or infrastructure resources. In the United States, FCC and NTIA are investing in the projects of spectrum sharing between Federal and non-Federal users as a means to increase spectrum capacity while increasing its efficiency as well \cite{NTIA-SpectrumSharing}. For example, T-Mobile USA is currently sharing its spectrum with federal agencies \cite{TmobFCC_SpectrumSharing}.

A major impediment for pervasive spectrum sharing is the providers' tendency to protect the bands that they procured with extensive licensing and operating costs. Current spectrum usage heavily follows competitive auctions~\cite{2010-moore-spectrum} which are healthy for balancing standardization trends~\cite{2002-desourdis-emerging}, but may hinder the potential gains that can be obtained via extensive sharing \cite{2011-jesuale-lights,2006-baumol-toward}. The traditional spectrum policies which are prevailing in the last few decades were developed with a motivation to minimize \emph{harmful interference} between the operators. However, the optimum spectrum policy is not to just eliminate the interference, rather to jointly optimize the total functionality, economic and societal benefits of all radio systems \cite{6732919}. Current broadband policies~\cite{national-broadband-plan} dictate that competitive auctions must remain intact, while simultaneously incorporating elements of spectrum sharing and new ways to manage spectrum usage. Such governance and policies welcoming the necessary economic tools for enabling heterogeneous spectrum sharing for \emph{the larger good} are imperative.

One promising approach to foster more sharing of the spectrum is via the use of regulatory power. In fact, the National Broadband Plan \cite{national-broadband-plan} recommends the wide-spread development of the concept of \emph{spectrum subsidy}, e.g., licensing of the D block for commercial use if \emph{public safety partnerships} are considered by the licensee.
In this paper, we investigate this idea of subsidizing the spectrum to the providers with lower costs in return of \emph{proof-of-sharing}. Thus, the providers will be offered discounted spectrum bands, potentially at different locations, but will be asked to cover users not subscribed to them so as to maintain their subsidy incentives from the government.
Recent studies suggest significant market and user welfare gains under such subsidization, e.g., data subsidy for offering minimal data plan to users for free \cite{2012-yu-guaranteeing} or spectrum pooling among providers to improve user experience \cite{2011-jesuale-lights}.
However, using game theory to analyze a subsidized spectrum market, that considers spatial and temporal provider-government relationships as well as dynamic components like roaming, and signal quality, is essentially unexplored. In this paper, we contribute a spectrum sharing framework that considers these aspects all at once in a noncooperative game.

The rest of the paper is organized as follows. Section~\ref{sec:related-work} summarizes the existing works on spectrum sharing in the literature. Section~\ref{sec:contributions} outlines the contributions of this paper. Section~\ref{sec:approach} presents a spectrum market between government, providers, and customers, and also highlights some key observations. In Section~\ref{sec:PerfectNashEquilibrium}, we describe this market as a noncooperative game, defining precisely the notion of perfect Nash equilibrium.  In Section~\ref{sec:AnalyticSol}, we formulate the multi-criteria maximization problem for the providers' part of the perfect equilibrium, in a simplified case of \emph{two providers and two regions}; in Section~\ref{sec:AnaResults}, we derive approximation to that equilibrium. There, we argue that in our specific case an ``entire" perfect equilibrium exists (including a regulatory strategy for the government). In Section~\ref{sec:result}, we analyze the results and provide insights into the subsidization model. There, we also show our simulations based on best response algorithm to converge to the equilibrium. Section~\ref{sec:summary} concludes the paper.

\section{Related Work}
\label{sec:related-work}
Regulatory practice has been part of the wireless market since its inception as the resource being shared was soon recognized to be very valuable. Governmental agencies establish regulatory policies to increase the efficiency of the spectrum market while preserving an acceptable level of fairness in sharing the precious spectrum. The concept of subsidization is one way to involve the government so as to incentivize various long-term goals in the market being regulated. Typically, subsidy regulations are used for markets serving ``the larger good'' (i.e., the benefit of the whole society) and involving a significant amount of cost producers cannot bare individually. Transportation \cite{1997-Karlaftis-subsidy}, agriculture \cite{1993-yabuuchi-urban} and telecommunications \cite{1998-parsons-cross} are examples of such markets where new initiatives require significant infrastructure investments, and hence the government's involvement is justified. However, subsidization is known to have negative effects on the efficiency of the market. Thus, \emph{performance-based} subsidization is preferred if at all possible.

Even though subsidization is heavily employed in several markets, the wireless market has not seen much of its usage beyond a few, limited scenarios. The most known subsidization in a wireless market is the subsidization of the expensive phones to the users by the provider \cite{2006-kelkar-over}, in that the user pays the phone's cost over a locked, termed contract. Unlocking the contract term either requires return of the phone or payment of a significant fee.

Only recently, Yu and Kim \cite{2012-yu-guaranteeing} focused on the concept of using subsidies for spectrum management. Their work analyzed price and quality of service (QoS) subsidy schemes to increase the utility of the consumers from the data plans. The goal is to increase the availability of wireless data plans to more users. In their model, the regulator offers a subsidized (i.e., less expensive) data plan to the end users with lower quality of service via subsidization to the providers. In our work, we consider subsidizing the spectrum to the providers and focus on implementing a subsidy system mainly between the providers and the government. We aim to make the subsidization seamless to the end users and aim to incentivize the providers to be more welcoming to the users subscribed to other providers. As a major difference, we focus on spectrum subsidy rather than data plan subsidy.

A related concept to the subsidization approach considered in this paper is inter-operator \emph{roaming}. Through roaming, a customer can make/receive calls and send/receive data while outside of the home network coverage, by using the infrastructure of another provider. This is enabled through reciprocal roaming charges among operators, which may vary, e.g., depending on the nationwide coverage supported by an operator through its infrastructure. The new roaming regulations in European Union (EU) to be effective from December 2015 will no longer allow the telecom providers to charge extra roaming costs to their customers \cite{EU_AbolishRoamingFee}. The goal is to facilitate competition in the roaming market and bring international roaming prices down to domestic rates, e.g., through diversity of market players, low barriers to market entry, and equal access to basic wholesale services. The analysis in~\cite{Fabrizi_TP2008} shows that, as long as certain coverage conditions are fulfilled, providers in an open system (where roaming is allowed) have strictly better revenues when compared to a closed system with no roaming. Benefits of inter-operator spectrum sharing and joint radio resource management (JRRM) have also been demonstrated in recent works~\cite{Giupponi_ICC2007,Chang_GLOBECOM2011,Berry_INFOCOM2013}. The inter-operator JRRM technique proposed in~\cite{Giupponi_ICC2007} allows subscribers to get service through other operators in case the home operator network is blocked, and results show that inter-operator roaming agreements improve both the network performance and the operators' revenue. 

In~\cite{Chang_GLOBECOM2011}, a roaming market model is introduced, which uses the roaming rate as an incentive for service providers to gain revenue, when they allow other users to access their network. Optimal roaming rate is derived to maximize the social welfare of all the service providers and licensed users. Our approach for subsidization, while also allows inter-operator sharing of spectrum similar to roaming, is fundamentally different than these earlier work, since it involves the government as the subsidy source to facilitate spectrum sharing.

Recently, dynamic spectrum sharing has been gaining interest where the service providers can buy additional channel bands to meet their customer demands or can sell their underutilized spectrum to gain an additional revenue. To this end, there are several works in the literature which are based on auctions for the secondary spectrum access \cite{6179566, 6657815, 4770127}. Typically, there exists a spectrum broker who divides the available spectrum into channels and allocates them to the competing providers using an auction mechanism. Determining a proper auction mechanism is of importance in this method. In \cite{6179566}, a generalized Branco's auction mechanism is introduced which aims at maximizing the profit to the seller, while in \cite{6657815}, a truthful auction mechanism is proposed which aims at maximizing the social welfare or the revenue of the seller. In \cite{4770127}, a knapsack-based auction mechanism is proposed for dynamically allocating the spectrum in the coordinated access band. Another set of works \cite{5173512, 6560443, 5332277} in the literature deals with the issues of spectrum pricing when the primary license holder wants to lease the spectrum to a secondary operator.

Even though dynamic spectrum sharing improves the spectrum utilization efficiency, there are few downsides that affect the network operators. Spectrum sharing agreements while provide the operators an alternative means of meeting their bandwidth demands, they may also encourage them to under-invest in their infrastructure. Such trade-offs have been investigated in~\cite{Berry_INFOCOM2013} considering two different sharing models. Furthermore, when a primary license holder leases his spectrum within a sub-region to a secondary operator, it results into a reduced spatial coverage of the primary operator's network, and also possible interference from the secondary operator. This trade-off is studied in \cite{5332277}. Our proposal can coexist with such dynamic sharing schemes since it can be orthogonal to the dynamic sharing deals among providers.

\section{Contributions}
\label{sec:contributions}
In the context of the other related work discussed in Section~\ref{sec:related-work}, the main contributions of this paper are to model a subsidization framework using an extensive form game, and to show how to find (perfect) equilibrium solutions for (a simple case of) the game. The players in the game are a finite number of service providers plus a single government player. [Customers are not formal players in our game as their actions are completely determined by the moves of the providers and the government.]  The government incentivizes the providers to give service to the foreign customers who are outside the coverage of their home provider. The providers each aim to maximize their profit from enrolling customers, by efficiently investing the subsidies received from the government.  A customer's utility depends upon the level of service obtained from its ''home provider". The government attempts to maximize social welfare, here measured as the sum of the customers' utilities.  

Initially, with preliminary observations of the game theoretic framework, we provide insights into the effects of subsidization on the service providers and the customers. However, due to the complexity of the problem, we cannot obtain closed form expressions for the equilibrium solutions. For this reason, we consider a simple, yet insightful case of \emph{two providers and two regions}. Here, we derive an approximation to part of the equilibrium solution using a heuristic approach. We check the accuracy of the approximation by comparing its results with the numerical equilibrium solutions. Since exact expressions to the equilibrium solutions could not be derived, the existence of Nash equilibrium could not be proved analytically. However, we argue that our analysis implies this existence at least for some particular parameter values.


\section{A Spectrum Market with Subsidy}
\label{sec:approach}
\begin{figure}[!t]
\centering
\includegraphics[width=9cm]{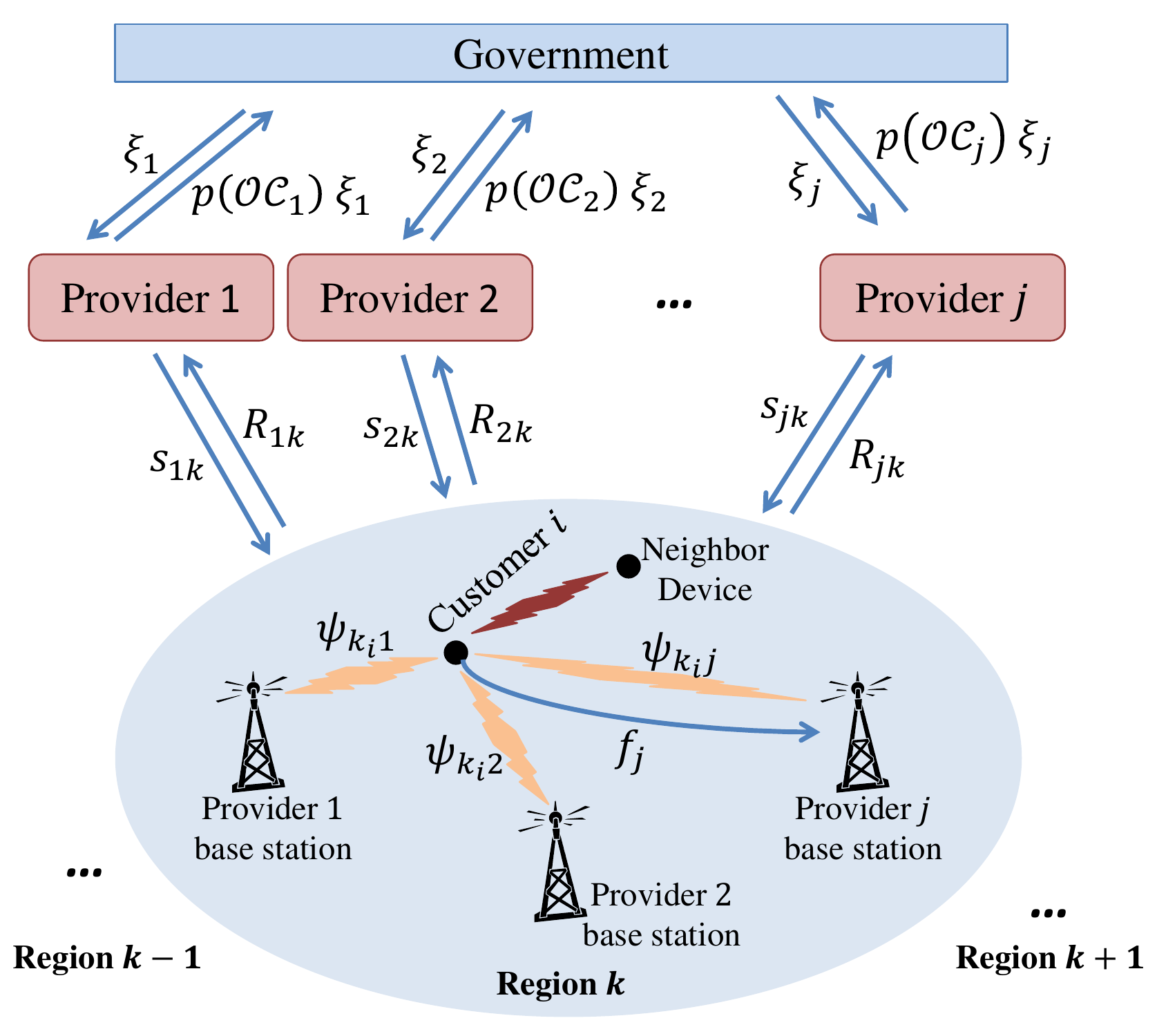}
\caption{Model for a subsidized spectrum market. In this model, government allocates subsidy $\xi_j$ to provider $j$, which in return can utilize a portion $s_{jk}$ of it to improve the service quality (which is captured by $\psi_{k_ij}$) in region $k$. Provider $j$ charges fee $f_j$ to each of its customers and as a result collects a revenue $R_{jk}$ from region $k$.}
\label{fig:subsidy-market}
\end{figure}
Even though there are several works in the literature on the role of government in market formations \cite{2011-shubik-theory, 2013-quint-barley}, the government's regulatory role in the radio spectrum market has not been explored well within a game-theoretic framework. We propose a spectrum market model with three types of agents as shown in Fig.~\ref{fig:subsidy-market}: customers, providers, and a single government player. A list of all the notations used in our model appears in Table~\ref{tab:symbols}. The customers are essentially the end-user devices that will ultimately engage in localized spectrum sharing markets. Let ${\cal I} = \{1,...,I\}$ be the set of wireless customers. These customers are spread out over a set of regions ${\cal K} = \{1,...,K\}$.

We assume there are $n_k$ customers in region $k$, $k = 1,...,K$. If $k_i$ is the region in which customer $i$ is located, we call $k_i$ its \emph{home region}.  Each customer $i$ makes $\beta_i$ calls in its home region (over the time period associated with the game); for simplicity we assume $\beta_i$ is equal to a constant $\beta$ for all $i$.  In addition, we assume the customers each make $\alpha$ calls outside their home region. To choose its main service provider, customer $i$ takes as given the ``intensity of service/signal'' $\psi_{k_i j}$ offered by each provider $j$ in region $k_i$, as well as the fee $f_j$ that each $j$ charges for service.  This is done probabilistically as described below.

The set of all providers is denoted by ${\cal J} = \{1,...,J\}$.  Providers operate in all regions.  The government gives each provider an allotment of bandwidth and of monetary subsidy, which they, in turn, allocate for their own use in each of the $K$ regions.  The amount of bandwidth and money spent in each region determines the intensity of service offered, which in turn helps determine who wins customer subscriptions. An important feature of this model is that the government motivates the providers to give service to customers who are outside of their home region (i.e., roaming) or simply far away from the base stations of their home providers. Foreign providers might be able to provide a higher intensity signal than the customers' home providers.  The government aims to motivate foreign providers by reducing each provider's subsidy if the provider does not service enough number of foreign customers who are away from their home providers.

We propose a two-stage, extensive form noncooperative game in which the government moves first, followed by the providers (all simultaneously).  Here, the customers are {\it not} formal players in the game because their actions are completely determined by the actions of the other players. Mathematically, we would like to solve for a perfect Nash equilibrium in the game as follows: a) Knowing what the customers will do as a function of $\left\{\psi_{k_i j}\right\}_{i,j}$ and $\left\{f_j\right\}^J_{j=1}$, we can solve for the optimal strategy for the providers; b) Knowing how the providers/customers behave, we can solve the government’s problem, which would be to maximize social welfare -- which in this case would be the total of customers'’ utility from consuming the wireless spectrum.
\begin{table}[!t]
\caption{Notations and Symbols}
\label{tab:symbols}
\centering
\begin{tabular}{|l|p{6.8cm}|}
\hline
\emph{Symbol} & \emph{Description} \\
\hline \hline
$I$       	& Total number of customers \\ \hline
$J$       	& Total number of providers \\ \hline
$K$       	& Total number of regions \\ \hline
$n_k$ 		& Number of customers in region $k$ \\ \hline
$\beta$ 	& Number of calls made by a customer in its home region \\ \hline
$\alpha$	& Number of outside calls a customer makes \\ \hline
$k_i$ 		& The region where customer $i$ is located \\ \hline
$f_j$ 		& The fee charged by provider $j$ to its customers \\ \hline
$\psi_{k j}$& The intensity/quality of provider $j$'s signal in region $k$ \\ \hline
$Q(.)$ 		& An increasing concave function used to calculate $\psi_{kj}$ -- parameterized with the investment, $s_{j k}$, and/or bandwidth, $b_{j k}$, in a region  \\ \hline
$u(.)$		& Customers' utility function -- parameterized with received signal intensity $\psi_{k j}$ \\ \hline
$U_{j}(k)$	& The utility obtained by choosing provider $j$ in region $k$ \\ \hline
$P_{j}(k)$	& The probability that provider $j$ is chosen in region $k$ \\ \hline
$R_{j k}$	& Provider $j$'s revenue collected from region $k$ \\ \hline
$s_{j k}$	& Portion of the subsidy amount provider $j$ designates/spends in region $k$ \\ \hline
$E_j$	      & Cash on hand for provider $j$, before obtaining any subsidy \\ \hline
$\xi_j$		& Subsidy amount provider $j$ receives from the government \\ \hline
$\xi$		& Total subsidy budget of the government \\ \hline
${\cal OC}_j$	& Number of outside calls provider $j$ serves \\ \hline
$p(.)$		& Penalty function determining the proportion of the subsidy to be returned to the government -- parameterized with ${\cal OC}_j$ \\ \hline
$\delta$ 	& Per-outside-call reward to the provider from the government \\ \hline
$\gamma$ 	& Scaling factor in the customer utility function\\ \hline
\end{tabular}
\end{table}

\subsection{Customer $i$'s Problem}
Each customer $i$ has the same increasing concave utility function $u$ for intensity of wireless service, expressed in units of money per call. Thus 
\begin{equation}
U_{j}(k_i) = \beta u (\psi_{k_i j} ) - f_j \label{eq:CustomerUtility}
\end{equation}
measures the utility a customer would get from local calls in its home region $k_i$, if the customer chose provider $j$ as its home provider. The customer then chooses provider $j$ ($j = 1,...,J$) with probability 
\begin{equation}
P_{j}(k_i) = { U_{j}(k_i) \over \sum_{j'=1}^J U_{j'}(k_i)}. \label{eq:ProviderPrblty}
\end{equation}
This probabilistic assignment of contracts (prizes) to providers (contestants) based on resources offered is taken from Contest Theory, and arises in many applications (see, e.g., \cite{2013-chowdhury-experimental}, \cite{2010-kovenock-conflicts}, or the excellent survey paper of \cite{2012-dechenau-survey}).

We note that the customers choose the providers based only upon offered service in their home regions, because their outside calls will always be covered (at no extra cost to them) as a result of the government's subsidization scheme.

\subsection{Provider $j$'s Problem}
Provider $j$ has at its disposal a total amount of cash $E_j + \xi_j$, where $E_j$ is the amount of cash on hand at the beginning of the game, and $\xi_j$ is the amount of subsidy it receives from the government.  They also are allocated bandwidth $B_j$ from the government.  Their problem is to allocate these resources among the regions, designating $s_{jk}$ to spend in region $k$ and bandwidth $b_{jk}$ to use in region $k$, $k = 1,...,K$.  The intensity of service that it offers in region $k$ is then
\begin{equation}
\psi_{kj} = Q(s_{jk}, b_{jk}),
\label{eq:intensity}
\end{equation}
where $Q$ is an exogenously given function (increasing and concave). In other words, the service intensity monotonically increases with the cash invested by the provider, as well as the utilized bandwidth.  Provider $j$ also sets the fee $f_j$ which all of its contracted customers in all regions must pay.

From the description of the customers' problem above, we see that provider $j$ will capture 
$n_k P_{j}(k)$
of the customers in region $k$, for a revenue of
\begin{equation}
R_{jk} = f_j n_k P_{j}(k),
\label{eq:revenue}
\end{equation}
where $P_j(k)$ is defined in (2).  Then their total revenue (across all regions) is $\sum_{k=1}^K R_{jk}$. 

Next, we note that the customers in region $\hat k$ make $\alpha n_{\hat k}$ calls outside of their home region. We assume
\begin{equation}
n_{\hat k k} = {\alpha n_{\hat k} n_k \over \sum_{k' \ne \hat k} n_{k'} }
 \label{eq:foreign-calls-region}
\end{equation}
of these occur in region $k$ ($k \ne \hat k$). These calls will be allocated probabilistically to the providers, according to the intensity levels offered by the providers in region $k$. The total number of such outside calls that provider $j$ serves is then
\begin{equation}
 {\cal OC}_j = \sum_{k=1}^K  \sum_{\hat k \ne k} n_{\hat k k} {u (\psi_{k j} )  \over \sum_{j'=1}^J u (\psi_{k j'} ) }.
 \label{eq:foreign-calls-provider}
\end{equation}
The provider is then assessed a penalty according to this number -- the lower the quantity ${\cal OC}_j$ is, the higher the penalty.  The penalty is expressed as a proportion of subsidy lost, and is incurred after the market.  So, assuming this penalty function is denoted $p(x)$, we can write down an optimization problem for provider $j$:
\begin{align}
\smash{\displaystyle\max_{\{s_{jk}\},\{b_{jk}\},f_j }} \hspace{3mm} \sum_{k=1}^K R_{jk} +& (1 - p({\cal OC}_j)) \xi_j - \sum_{k=1}^K \hspace{-1mm} s_{jk} \label{eq:provider-prob}\\
{\rm such~that}, \hspace{3mm} \sum_{k=1}^K s_{jk} \le & E_j + \xi_j, \label{eq:provider-cons1} \\
\sum_{k=1}^K b_{jk} =& B_j, \label{eq:provider-cons2} \\
\psi_{kj} =& Q(s_{jk}, b_{jk}), \label{eq:provider-cons3} \\
s_{jk}, b_{jk}, f_j \ge& 0,\ \forall~j,k,\label{eq:provider-cons4}
\end{align}
where the first term in the maximization is the total revenue, the second term is the leftover subsidy money after the penalty is deducted due to fewer-than-required sharing of the provider's spectrum, and the last term is the total amount of expenses the provider uses from the subsidy it received from the government.

\subsection{Government's Problem}
\label{sec:GovtProb}
Finally, the government's decision variables are $\xi_j$ and $B_j$, $j=1,...,J$.  It also can designate the function $p(x)$ described above.  Its objective is to maximize social welfare, here defined as the customers' total utility from wireless service, which is
\begin{align}
\smash{\displaystyle\max_{\{\xi_j, B_j\}_{j=1}^J}} & \left[\sum_{i=1}^I \beta u(\psi_{k_i j(i)}) \right. \nonumber\\
& \left. + \sum_{j=1}^J \sum_{k=1}^K \left( \sum_{\hat k \ne k} {n_{\hat k k} \ u(\psi_{k j} )  \over \sum_{j'=1}^J u (\psi_{k j'} ) } u(\psi_{kj}) \right) \right] \label{eq:govern-prob}\\
\rm{s.t.,}\ \ \ \ & \sum_{j=1}^J B_j = B, \label{eq:govern-cons1}\\
& \sum_{j=1}^j \xi_j \le \xi, \label{eq:govern-cons2}\\
& B_j, \xi_j \ge 0\ \forall j. \label{eq:govern-cons3}
\end{align}
Here, the first term is the utility that customers acquire in their home regions, while the second term is from calls in outside regions.  The notation $j(i)$ is the home provider to which customer $i$ is assigned.  The quantities $B$ and $\xi$ represent the total amounts of bandwidth and subsidy that the government has at its disposal to allocate.

\subsection{Observations and Discussion}
The model we have just outlined, while not perfectly covering all details and aspects, provides a good framework to consider the viability of subsidizing spectrum to providers in return for increased spectrum sharing. The providers constitute a key player to scrutinize in such a spectrum market model. Ultimately, the goal of this market model is to motivate the providers to share their spectrum among each other to attain the larger good of serving more customers with better quality access links.

Provider $j$'s problem (\ref{eq:provider-prob}) offers several insights. The \emph{first term} in (\ref{eq:provider-prob}) is the fees provider $j$ collects from its home customers, who are subscribed to provider $j$. Provider $j$ has to serve these customers by default. But when these customers travel to other regions, they may be able to find another provider that offers a better quality signal at that location. In that case, those customers would be considered as making an outside call, and thus contributing to the sharing of the spectrum among the providers.

The \emph{second term} in (\ref{eq:provider-prob}) describes the amount of subsidy money left to provider $j$. This leftover amount is dependent on the subsidy amount, $\xi_j$, given by the government to provider $j$, and more importantly, the number of outside calls, ${\cal OC}_j$, served by provider $j$ during the subsidy term. If ${\cal OC}_j$ is not large enough, provider $j$ may have to return all of the subsidy back to the government. At the other extreme, provider $j$ may keep all of the subsidy amount, $\xi_j$, if it did serve sufficient number of foreign customers.

The \emph{third term} in (\ref{eq:provider-prob}) quantifies the sum of investments provider $j$ makes from its subsidy. Provider $j$ can choose which regions to invest more. It can further decide whether to spend all of the subsidy it received, $\xi_j$, from the government. In this model, the provider takes some risk by spending the subsidy money. It needs to wisely decide on which regions to invest. The ideal situation for the provider is to invest into those regions in which more outside calls become possible to serve. This will motivate the providers to invest into more congested areas and thus serve the larger goal of improving user-perceived link quality.

\vspace{2mm} \noindent \underline{\textbf{Observation 1:}}
\emph{Providers cannot be hurt by the subsidization option.}

\vspace{1mm} \noindent
A provider (say $j$) does not have to take the subsidization option if it will not generate any profits. If $j$ does not participate in the subsidization plan, then the government cannot penalize it for not serving the required number of outside calls and therefore the revenue generated from the provider's own investment $E_j$ will remain intact.

\vspace{2mm} \noindent \underline{\textbf{Observation 2:}}
\emph{Providers can easily be motivated into a subsidized market.}

\vspace{1mm} \noindent
The government can offer a large $\xi_j$ and a conservative (i.e., not heavily penalizing) penalty function to attract the providers into the subsidization option. Since, from Observation 1, there is really not much of risk from the subsidization option, the tipping point for providers to sign into subsidization contracts will not be high. A relatively high $\xi_j$ will promise a positive return from the subsidization.

\vspace{2mm} \noindent \underline{\textbf{Observation 3:}}
\emph{Providers will be motivated to invest in a non-overlapping manner and collectively cover a larger area.}

\vspace{1mm} \noindent
The number of outside calls ${\cal OC}_j$ in (\ref{eq:provider-prob}) depends on the amount of overlap between the network coverage areas of different providers. Operators, by not duplicating their infrastructure in the same areas, can minimize their investment cost $s_{jk}$, while at the same time minimizing the penalty cost $(1-p({\cal OC}_j)) \xi_j$ charged by the government in (\ref{eq:provider-prob}), yielding high revenues. This observation agrees with the conclusions in~\cite{Fabrizi_TP2008}, in which the analysis suggests that minimization of the duplication of network infrastructure by different providers yields higher provider revenues.

\vspace{2mm} \noindent \underline{\textbf{Observation 4:}}
\emph{As the ratio ${\alpha \over \beta}$ increases, subsidization will be more beneficial, particularly to those providers smaller in size.}

\vspace{1mm} \noindent
Suppose $\alpha$ increases (and $\beta$ remains the same, so that the ratio  ${\alpha \over \beta}$ increases).  Then from (5) and (6), it follows that ${\cal OC}_j$ will also increase.  This will in turn increase the contribution of the second term in (\ref{eq:provider-prob}) to provider $j$'s revenue. An interesting situation happens for smaller providers. Small providers serve small regions, so their customers are more apt to leave their home regions.  Hence they make relatively more outside calls than home calls.  Hence the contribution of the second and third term of (\ref{eq:provider-prob}) into their revenue total could be more than the first term's contribution.

\vspace{2mm} \noindent \underline{\textbf{Observation 5:}}
\emph{Small providers will be more able to compete with large providers as the ratio $\alpha \over \beta$ increases.}

\vspace{1mm} \noindent
This follows from Observation 4. A new note to make here is that small providers with smaller infrastructure will be able to compete against providers with large infrastructure as long as $\alpha$ is relatively more than $\beta$.  They will be able to do so by swiftly investing into congested spots where larger providers cannot reach well.

\vspace{2mm} \noindent \underline{\textbf{Observation 6:}}
\emph{Monopolization will be avoided as long as the ratio $\alpha \over \beta$ is high.}

\vspace{1mm} \noindent
This follows from Observations 4 and 5. As we've seen, under this scenario smaller providers have the advantage. Thus, there will be no incentive to monopolize.

\section{Perfect Nash Equilibrium}
\label{sec:PerfectNashEquilibrium}
As stated previously, our game is one where the government moves first followed by the providers. This makes sense intuitively, as it is natural that the government ``makes the rules" and private companies then react to them.  The government strategy is to set the subsidy amounts $\xi_j$ and the bandwidths $B_j$ granted to each provider $j$.  After this, each provider simultaneously tries to maximize its utility by optimizing its strategic variables: subsidy amounts spent in each region $s_{jk}$, bandwidth allocated in each region $b_{jk}$ and fee charged to customers $f_j$. Game theoretically, the ``game tree" has two types of subtree:  the whole game itself, and also any subtree occurring after the government makes its initial move.

In game theory, a \emph{Nash equilibrium} is a set of strategies, one for each player, with the property that:  given the other players stick to their assigned strategies, no player can gain by unilaterally deviating from their own strategy. A \emph{perfect Nash equilibrium} is a special type of Nash equilibrium which has this property not only with regard to the whole game, but also if restricted (in the natural way) to any subtree of the game.  

In our game, a strategy for the government can be represented as $g = \{\xi_j, B_j\}_{j\in\mathcal{J}}$.  Denote the set of all such strategies by $G$.  Next, a strategy for provider $j \in \mathcal{J}$ can be represented by $a_j(g) = \{s_{jk}(g), b_{jk}(g), f_j(g)\}_{k\in\mathcal{K}}$. Note that the strategies for the providers have $g$ as an argument, as the providers move after the government and so can base their moves according to what the government does.  Finally, denote the strategies of all the providers collectively by the vector $\boldsymbol{a}(g) = \{a_j(g)\}_{j\in\mathcal{J}}$.

Given the above notations, the definition of perfect Nash equilibrium reduces to:
\begin{definition}
A set of strategies  $\left\{\boldsymbol{a}^*,g^*\right\}$ constitutes a \emph{perfect Nash equilibrium} of this game, if and only if, it satisfies the following set of inequalities:
\begin{align}
&U_{\rm p}^j\left(a_j^*(g), \boldsymbol{a}_{-j}^*(g), g\right) \nonumber\\
&\ \ \ \ \ \ \ \ \geq U_{\rm p}^j\left(a_j(g), \boldsymbol{a}_{-j}^*(g), g\right) , \ \forall a_j,\ j\in\mathcal{J}, \ {\rm{and}}\ g\in G\nonumber\\
& \mbox{and} \ \ U_{\rm G}\left(g^*, \boldsymbol{a}^*(g^*)\right) \geq U_{\rm G}\left(g, \boldsymbol{a}^*(g)\right)\ \forall g\in G,
\end{align}
where, $\boldsymbol{a}_{-j}^* = \{a_i\}_{i\in \mathcal{J}, i\ne j}$, is the vector of strategies of all the providers other than $j$. The function $U_{\rm p}^j$ is the provider~$j$'s utility given by the expression in maximization problem of \eqref{eq:provider-prob}. The function $U_{\rm G}$ is the government's utility given by the expression in maximization problem of \eqref{eq:govern-prob}.
\end{definition}

In general, for our game the conditions (16) are too difficult to solve analytically, or even computationally.  The best we can do is to consider a simplification of the game in which a) there are only two regions and two providers; b) the bandwidth variable has been taken out of the model; and c) there is a linear penalty function.  Assuming particular parameter values, we can computationally solve for the providers' equilibrium strategies as a function of $g$; from this we argue that a perfect Nash equilibrium must exist.  See Section~\ref{sec:NashEquExistance}.


\section{A Simplified Spectrum Market}
\label{sec:AnalyticSol}
In this section, we simplify the market model in Section~\ref{sec:approach}. In particular, we make the following simplifications:
\begin{itemize}
\item{} We assume a linear penalty function $p({\cal OC}_j)$, for which $p(0) = 1$ (if a provider serves no foreign customers, it loses all of its subsidy). It implies that the $(1-p({\cal OC}_j)) \xi_j$ term in (7) can be recast as a ``reward function" $\delta {\cal OC}_j$, where $\delta$ is a per-outside-call reward that the government gives to providers.
\item {} By eliminating the explicit handling of bandwidth and assuming that it is implicit in the intensity function that expresses the received signal quality on an access link. Hence, \eqref{eq:intensity} reduces to $\psi_{k j} = Q(s_{k j})$.
\item {} We assume a concave (square root) utility function $u$, linear signal quality function $Q$, and set the initial cash holdings $E_j$ equal to zero.
\end{itemize}
We then consider a particular case with two-regions and two-providers to obtain closed form expressions for the providers' part of the equilibrium solution (subsidy $s^*_{jk}$ and fee $f^*_j$), as a function of the government's strategy $\xi$.

\subsection{A Linear Penalty Function}
\label{sec:Lin_Penalty_Func}
\begin{figure}[htp]
\begin{center}
\includegraphics[width = 1.7 in]{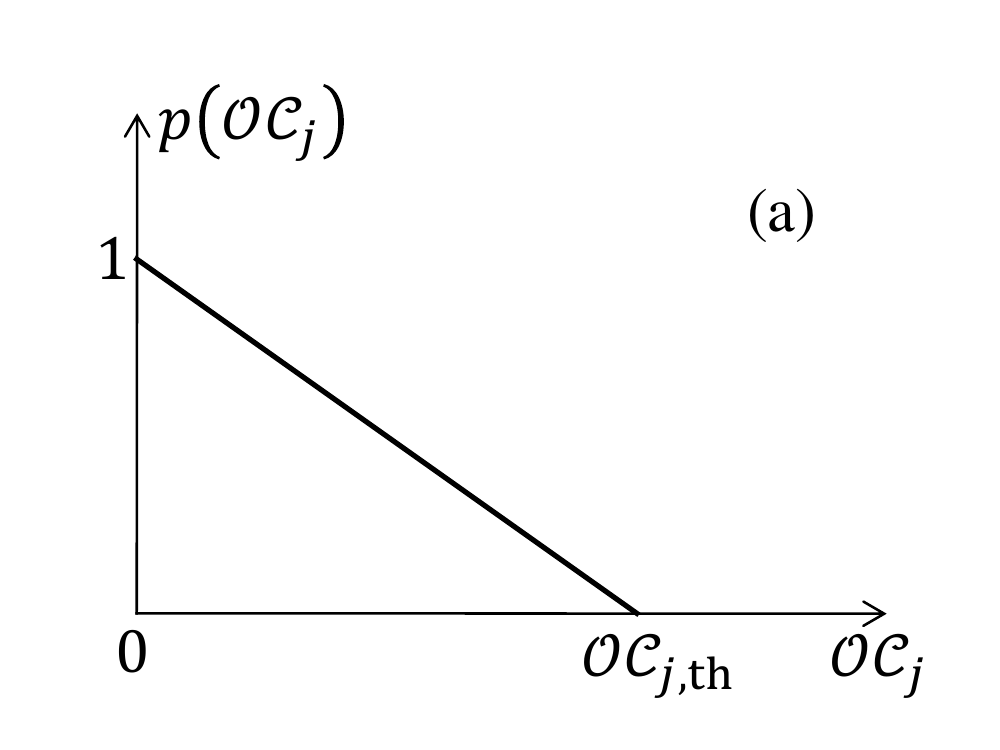}
\includegraphics[width = 1.7 in]{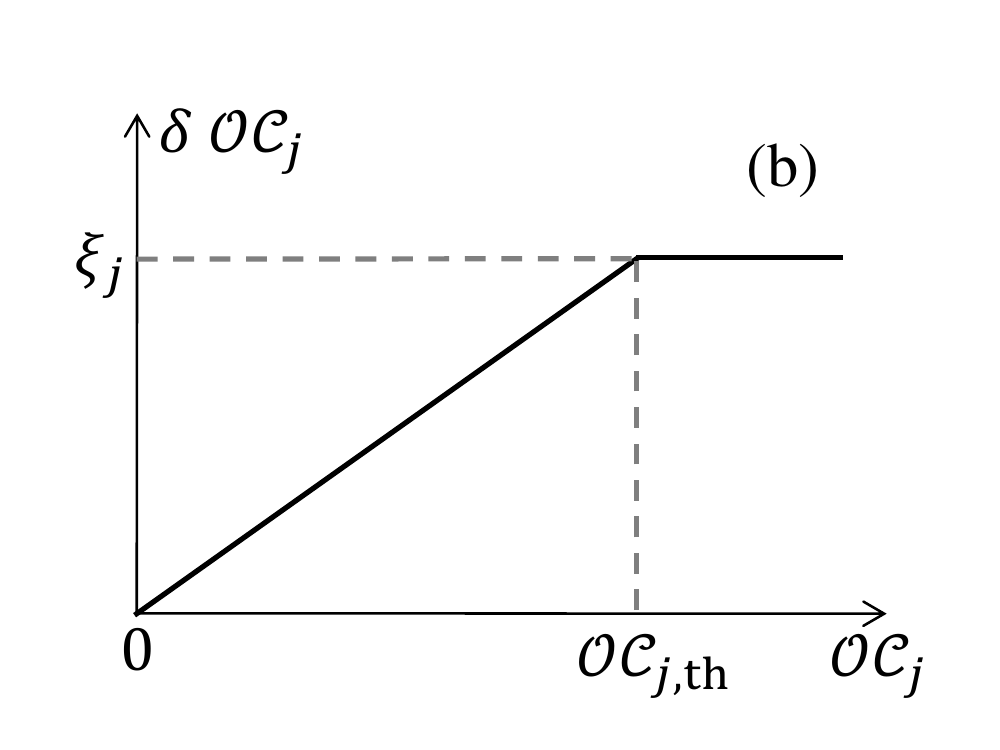}
\end{center}
\caption{Illustration of (a) penalty function; (b) reward function.}
\label{fig:PenaltyFunction}
\end{figure}

The linear penalty function described above is illustrated in Fig.~\ref{fig:PenaltyFunction}(a) which is a linearly decreasing function of $\mathcal{OC}_j$. If provider $j$ serves a minimum of $\mathcal{OC}_{j,{\rm th}}$ number of outside calls, then the penalty to the provider will be zero.
On the other hand, the reward function illustrated in Fig.~\ref{fig:PenaltyFunction}(b) is an increasing function of $\mathcal{OC}_j$. If provider $j$ serves a minimum of $\mathcal{OC}_{j,{\rm th}}$ number of outside calls, then it would receive complete subsidy amount $\xi_j$, which is equivalent to zero penalty. Here, the slope of reward function is the per-outside-call reward $\delta$ that is specified by the government. Based on the available budget $\xi$ and the expected number of outside calls $\alpha I$, government can determine the per-outside-call reward as
\begin{align}
\delta = \frac{\xi}{\alpha I} = \frac{\sum_{j=1}^J \xi_j}{\alpha\sum_{k=1}^{K}n_k}. \label{eq:DeltaExpr}
\end{align}
Then, provider $j$ has to serve $\mathcal{OC}_{j,{\rm th}} = \xi_j/\delta$ number of outside calls to receive the maximum reward $\xi_j$.

\subsection{Bandwidth Implicit in the Intensity Function}
As stated above, we assume $\psi_{k j} = Q(s_{k j})$. Using this, the linear form for the penalty function, and the expressions \eqref{eq:ProviderPrblty}--\eqref{eq:foreign-calls-provider} and \eqref{eq:DeltaExpr}, we may rewrite provider $j$'s problem as follows:
\begin{align}
\smash{\displaystyle\max_{s_{j1},s_{j2},...,s_{jK},f_j }} & \sum_{k=1}^K f_j n_k {\beta u (Q(s_{jk}) ) - f_j \over \sum_{j'=1}^J \beta u (Q(s_{j' k})) - f_{j'} } \nonumber\\
& \hspace{-1.9cm} + \frac{\xi}{I} \sum_{k=1}^K  \sum_{\hat{k} \ne k} {n_{\hat k}\ n_{k} \over \sum_{k' \ne \hat k} n_{k'} } {u (Q(s_{jk}))  \over \sum_{j'=1}^J u (Q(s_{j'k})) } - \sum_{k=1}^K s_{jk} \label{eq:provider-simple-1}\\
\rm{s.t.,}\ \ \ \ & \sum_{k=1}^K s_{jk} \le E_j + \xi_j, \label{eq:provider-simple-cons1} \\
& s_{jk} \ge 0, k=1...K, \label{eq:provider-simple-cons2} \\
& f_{j} \ge 0. \label{eq:provider-simple-cons3}
\end{align}
Here, it is worthwhile to note that \eqref{eq:provider-simple-1} is independent of the parameter $\alpha$ due to the use of linear penalty function and the assumption that each customer makes same number of outside calls (i.e., $\alpha$).

\subsection{Further Simplifying Assumptions}
\label{sec:Further_Simpl_Assumptions}
We further make the following simplifying assumptions for our market:

\begin{itemize}
\item Concave customer utility function of the form: $u(x) = \gamma \sqrt{x}$, where $\gamma \ll 1$ is a scaling factor.
\item A linear signal quality function: $Q(s_{j k}) = s_{j k}$.
\item No initial cash on hand for the providers, i.e. $E_j = 0$ for all $j$.
\end{itemize}
In general, a concave (square-root) utility in the first assumption is considered realistic enough to capture the diminishing returns behavior of received quality. The second assumption on linear signal quality function may be considered optimistic, since the signal quality will also behave in a diminishing returns manner as the investment on the infrastructure increases. However, since $Q(\cdot)$ always feeds into $u(\cdot)$ in our formulation, the $u(Q(\cdot))$ will still behave according to diminishing returns.

\subsection{Two Regions and Two Providers}
\label{sec:TwoRegionsTwoProviders}

To make our model more understandable, we now rewrite the providers' problems for the case with two providers and two regions. All of the assumptions in Sections~\ref{sec:Lin_Penalty_Func} through \ref{sec:Further_Simpl_Assumptions} are incorporated here:
\subsubsection{Provider 1's Problem}
By applying the aforementioned simplifying assumptions into \eqref{eq:provider-simple-1}, the objective function of provider~1 can be expressed as:
\begin{align}
\smash{\max_{s_{11}, s_{12}, f_1}}\ & f_1 n_1 \frac{\gamma \beta \sqrt{s_{11}} - f_1}{\gamma \beta \sqrt{s_{11}} - f_1 + \gamma \beta \sqrt{s_{21}} - f_2} \nonumber\\
& + f_1 n_2 \frac{\gamma \beta \sqrt{s_{12}} - f_1}{\gamma \beta \sqrt{s_{12}} - f_1 + \gamma \beta \sqrt{s_{22}} - f_2} \nonumber \\
& + \frac{\xi}{I} \left(n_1 \frac{\sqrt{s_{12}}}{\sqrt{s_{12}} + \sqrt{s_{22}}} + n_2 \frac{\sqrt{s_{11}}}{\sqrt{s_{11}} + \sqrt{s_{21}}}\right) \nonumber\\
& - s_{11} - s_{12}, \label{eq:provider1} \\
\rm{s.~t.,} \ \ \ \ & s_{11} + s_{12} \le \xi_1, \label{eq:provider1-cons1} \\
& s_{11}, s_{12}, f_1 \ge 0. \label{eq:provider1-cons2}
\end{align}
We take the first-order conditions for the system \eqref{eq:provider1}--\eqref{eq:provider1-cons2} as follows:\\
w.r.t. $f_1$:
\begin{align}
& \frac{\splitfrac{n_1 (\gamma\beta\sqrt{s_{11}} - 2 f_1) (\gamma \beta \sqrt{s_{11}} - f_1 + \gamma \beta \sqrt{s_{21}} - f_2)} {+ f_1 n_1 (\gamma \beta\sqrt{s_{11}} - f_1)}}{(\gamma \beta \sqrt{s_{11}} - f_1 + \gamma \beta \sqrt{s_{21}} - f_2)^2} \nonumber \\
& + \frac{\splitfrac{n_2 (\gamma \beta \sqrt{s_{12}} - 2 f_1)(\gamma \beta \sqrt{s_{22}} - f_2 + \gamma \beta \sqrt{s_{12}} - f_1)} {+ f_1 n_2 (\gamma \beta \sqrt{s_{12}} - f_1)}}{(\gamma \beta \sqrt{s_{22}} - f_2 + \gamma \beta \sqrt{s_{12}} - f_1)^2}\nonumber\\
&\hspace{6cm}=0,  \label{eq:provider1-foc1}
\end{align}
w.r.t. $s_{11}$:
\begin{align}
&\frac{\beta f_1 n_1 \gamma \left(\gamma \beta \sqrt{s_{21}} - f_2 \right)}{2\sqrt{s_{11}} \left(\gamma \beta \sqrt{s_{11}} - f_1 + \gamma \beta \sqrt{s_{21}} - f_2 \right)^2}\nonumber\\
&+ \frac{n_2 \xi \sqrt{s_{21}}}{2 I \left( \sqrt{s_{21}} + \sqrt{s_{11}} \right)^2 \sqrt{s_{11}}} - 1 - \lambda_1 = 0, \label{eq:provider1-foc2a}
\end{align}
w.r.t. $s_{12}$:
\begin{align}
&\frac{\beta f_1 n_2 \gamma \left(\gamma \beta \sqrt{s_{22}} - f_2 \right)}{2 \sqrt{s_{12}} \left(\gamma \beta \sqrt{s_{12}} - f_1 + \gamma \beta \sqrt{s_{22}} - f_2 \right)^2}\nonumber\\
&+ \frac{n_1 \xi \sqrt{s_{22}}}{2 I \left( \sqrt{s_{12}} + \sqrt{s_{22}} \right)^2 \sqrt{s_{12}}} - 1 - \lambda_1 = 0, \label{eq:provider1-foc2b}
\end{align}
and
\begin{align}
\lambda_1 \ge 0, ~ s_{11} + s_{12} = \xi_1 \hspace{2mm}\rm{~or}\hspace{2mm} \lambda_1 = 0, ~ s_{11} + s_{12} \le \xi_1. \label{eq:provider1-foc3}
\end{align}
Here, $\lambda_1$ is the Lagrangian multiplier for the constraint \eqref{eq:provider1-cons1} on the total amount of subsidization the government offers to providers.

\subsubsection{Provider 2's Problem}
The objective function of provider~2 can be expressed as
\begin{align}
\smash{\max_{s_{21}, s_{22}, f_2}}\ & f_2 n_1 \frac{\gamma \beta \sqrt{s_{21}} - f_2}{\gamma \beta \sqrt{s_{11}} - f_1 + \gamma \beta \sqrt{s_{21}} - f_2}\nonumber\\
& + f_2 n_2 \frac{\gamma \beta \sqrt{s_{22}} - f_2}{\gamma \beta \sqrt{s_{12}} - f_1 + \gamma \beta \sqrt{s_{22}} - f_2} \nonumber \\
& + \frac{\xi}{I} \left(n_1 \frac{\sqrt{s_{22}}}{\sqrt{s_{12}} + \sqrt{s_{22}}} + n_2 \frac{\sqrt{s_{21}}}{\sqrt{s_{11}} + \sqrt{s_{21}}}\right) \nonumber\\
& - s_{21} - s_{22}, \label{eq:provider2} \\
\rm{s.~t.,} \ \ \ \ & s_{21} + s_{22} \le \xi_2, \hspace{5mm} \mbox{and} \hspace{5mm} s_{21}, s_{22}, f_2 \ge 0. \label{eq:provider2-cons1}
\end{align}
Similar to Provider 1, Provider 2 will have to optimize w.r.t. $s_{21}$, $s_{22}$, and $f_2$. So, the first-order conditions will be:\\
w.r.t. $f_2$:
\begin{align}
& \frac{\splitfrac{n_1 (\gamma \beta \sqrt{s_{21}} - 2 f_2) (\gamma \beta \sqrt{s_{11}} - f_1 + \gamma \beta \sqrt{s_{21}} - f_2)} {+ f_2 n_1 (\gamma \beta \sqrt{s_{21}} - f_2)}}{(\gamma \beta \sqrt{s_{11}} - f_1 + \gamma \beta \sqrt{s_{21}} - f_2)^2} \nonumber \\
& + \frac{\splitfrac{n_2 (\gamma \beta \sqrt{s_{22}} - 2 f_2)(\gamma \beta \sqrt{s_{12}} - f_1 + \gamma \beta \sqrt{s_{22}} - f_2)} {+ f_2 n_2 (\gamma \beta \sqrt{s_{22}} - f_2)}}{(\gamma \beta \sqrt{s_{12}} - f_1 + \gamma \beta \sqrt{s_{22}} - f_2)^2} \nonumber\\
& \hspace{6cm} = 0, \label{eq:provider2-foc1}
\end{align}
w.r.t. $s_{21}$:
\begin{align}
&\frac{\beta f_2 n_1 \gamma \left(\gamma \beta \sqrt{s_{11}} - f_1 \right)}{2 \sqrt{s_{21}} \left(\gamma \beta \sqrt{s_{11}} - f_1 + \gamma \beta \sqrt{s_{21}} - f_2 \right)^2}\nonumber\\
&+ \frac{n_2 \xi \sqrt{s_{11}}}{2 I \left( \sqrt{s_{21}} + \sqrt{s_{11}} \right)^2 \sqrt{s_{21}}} - 1 - \lambda_2 = 0, \label{eq:provider2-foc2a}
\end{align}
w.r.t. $s_{22}$:
\begin{align}
&\frac{\beta f_2 n_2 \gamma \left(\gamma \beta \sqrt{s_{12}} - f_1 \right)}{2 \sqrt{s_{22}} \left(\gamma \beta \sqrt{s_{12}} - f_1 + \gamma \beta \sqrt{s_{22}} - f_2 \right)^2}\nonumber\\
&+ \frac{n_1 \xi \sqrt{s_{12}}}{2 I \left( \sqrt{s_{12}} + \sqrt{s_{22}} \right)^2 \sqrt{s_{22}}} - 1 - \lambda_2 = 0, \label{eq:provider2-foc2b}
\end{align}
and
\begin{align}
\lambda_2 \ge 0, ~ s_{21} + s_{22} = \xi_2 \hspace{2mm}\rm{~or}\hspace{2mm} \lambda_2 = 0, ~ s_{21} + s_{22} \le \xi_2, \label{eq:provider2-foc3}
\end{align}
where $\lambda_2$ is the Lagrangian multiplier.

\section{Optimal Policies and Existence of Perfect Nash Equilibrium}
\label{sec:AnaResults}
\subsection{Optimum Subsidy Amounts Spent ($s^*_{jk}$)}
Looking at the formulations of the providers' problems described in \eqref{eq:provider1}--\eqref{eq:provider2-foc3}, it is clear that the problem in hand is a multi-criteria maximization problem with \eqref{eq:provider1} and \eqref{eq:provider2} being the two simultaneous objective functions. The parameters to be optimized are the fees $f_1, f_2,$ and the subsidy amounts $s_{11}, s_{12}, s_{21},s_{22}$, which are functions of the input parameters $\beta, n_1, n_2, \xi_1$ and $\xi_2$. We wish to solve the eight simultaneous equations formed by the first order conditions in \eqref{eq:provider1-foc1}--\eqref{eq:provider1-foc3} and \eqref{eq:provider2-foc1}--\eqref{eq:provider2-foc3}, to obtain closed form expressions for the optimum parameters. Note that each parameter to be optimized depends on the other optimization parameters. Due to high number of unknown variables and the involved complexity while solving these simultaneous equations (such as solving polynomials of 7th order or higher), it is impractical to derive the expressions for optimum parameters in closed form. Hence, we derive approximations for the optimum fee and subsidy parameters by following a heuristic approach. In this approach, we numerically solve the simultaneous equations of \eqref{eq:provider1-foc1}--\eqref{eq:provider1-foc3} and \eqref{eq:provider2-foc1}--\eqref{eq:provider2-foc3}, and generate the plots of optimum fees to be charged to the customers, and optimum subsidy amounts to be spent by the providers in each region, for a wide range of input parameters. By studying these plots, we develop expressions for the optimum parameters by understanding the relationships of these parameters with each of the input parameters. The following important relationships were noted by studying the numerical plots.
\begin{enumerate}
\item The subsidy amounts $s^*_{11}$ and $s^*_{12}$ spent by provider 1 in regions 1 and 2 respectively, are proportional to the subsidy $\xi_1$ granted to provider 1 by the government and are independent of $\xi_2$. Similarly, $s^*_{21}$ and $s^*_{22}$ are proportional to $\xi_2$ and independent of $\xi_1$.
\item The providers each spend the entire amount of subsidy given to them by the government. That is, $s^*_{11}+s^*_{12}=\xi_1$, and $s^*_{21}+s^*_{22}=\xi_2$.
\item The $s^*_{jk}$'s vary non-linearly with respect to $n_1$ and $n_2$. Also, when $n_1 = n_2$, we have $s^*_{11} = s^*_{12}$ and $s^*_{21} = s^*_{22}$.
\end{enumerate}
The approximations for the optimum subsidy amounts that satisfy the above mentioned points were derived via a heuristic approach as
\begin{align}
s_{11}^* &\approx \frac{\xi_1}{2} \left(\frac{n_1}{n_1+n_2}+\frac{1}{2}\right),\label{s_11_approx}\\
s_{12}^* &\approx \frac{\xi_1}{2} \left(\frac{n_2}{n_1+n_2}+\frac{1}{2}\right),\label{s_12_approx}\\
s_{21}^* &\approx \frac{\xi_2}{2} \left(\frac{n_1}{n_1+n_2}+\frac{1}{2}\right),\label{s_21_approx}\\
s_{22}^* &\approx \frac{\xi_2}{2} \left(\frac{n_2}{n_1+n_2}+\frac{1}{2}\right).\label{s_22_approx}
\end{align}
The accuracy of these expressions will be verified in Section~\ref{sec:result} through comparisons with the numerical results.

\subsection{Optimum Fee ($f^*_j$)}
\label{subsec:Fj}
Similar to the approach that was used to find the optimum subsidy amounts to be spent by the providers, a heuristic approach is used to find the optimum fees to be charged by providers to the customers. The numerical solutions of the optimum fees $f^*_1$ and $f^*_2$ revealed that the optimum fees have negligible dependence on $n_1$ and $n_2$, as will be shown in Section~\ref{sec:result}. Hence, without loss of accuracy we make an assumption that the number of users in the two regions are the same, that is, $n_1 = n_2 = n$. Then, the providers' objective functions can be simplified which will allow us to mathematically derive the closed form expressions for the optimum fee.
\subsubsection*{Theorem 1}
In the simplified \emph{two-region two-provider} subsidization model, if the number of users in the two regions are assumed to be same, i.e., $n_1=n_2=n$, then the optimum fee $f_1^*$ and $f_2^*$ set by the two providers can be derived as
\begin{align}
f^*_1 =& \frac{\gamma \beta(7\sqrt{s^*_1}+4\sqrt{s^*_2})}{9}\nonumber\\
&-2\sqrt{\frac{A}{3}}\cos\left(\frac{1}{3}\cos^{-1}\sqrt{\frac{27B^2}{4A^3}}+\frac{\pi}{3}\right),\label{eq:f1}\\
f^*_2 =& \frac{\gamma \beta(7\sqrt{s^*_2}+4\sqrt{s^*_1})}{9}\nonumber\\
&-2\sqrt{\frac{C}{3}}\cos\left(\frac{1}{3}\cos^{-1}\sqrt{\frac{27D^2}{4C^3}}+\frac{\pi}{3}\right),\label{eq:f2}
\end{align}
where
\begin{align}
A &= \frac{4 \gamma^2 \beta^2\left[9\sqrt{s^*_1 s^*_2}+(\sqrt{s^*_1}-2\sqrt{s^*_2})^2\right]}{27},\\
B &= \frac{\gamma^3\beta^3}{27^2}\left(\splitfrac{16 s^*_1 \sqrt{s^*_1} - 240 s^*_2 \sqrt{s^*_1}} {- 123 s^*_1 \sqrt{s^*_2} - 128 s^*_2\sqrt{s^*_2}}\right),\\
C &= \frac{4 \gamma^2 \beta^2\left[9\sqrt{s^*_2 s^*_1}+(\sqrt{s^*_2}-2\sqrt{s^*_1})^2\right]}{27},\\
D &= \frac{\gamma^3 \beta^3}{27^2}\left(\splitfrac{16 s^*_2 \sqrt{s^*_2} - 240 s^*_1 \sqrt{s^*_2}} {- 123 s^*_2 \sqrt{s^*_1} - 128 s^*_1\sqrt{s^*_1}}\right),\\
s^*_1 &= \xi_1/2 {\rm\ \ \ \ \ and \ \ \ \ } s^*_2 = \xi_2/2. \label{eq:s1_s2}
\end{align}
\subsubsection*{Proof}
See Appendix. $\Box$

\subsection{Existence of Perfect Nash Equilibrium}
\label{sec:NashEquExistance}
In our simplified game, we now argue that the previous results imply the existence of a perfect Nash equilibrium (see Section~\ref{sec:PerfectNashEquilibrium}).  In particular, from \eqref{s_11_approx}--\eqref{eq:s1_s2} (and from the numerical results in Figs.~\ref{fig:Obj_sjk_vs_xi1_xi2} and \ref{fig:Fx_vs_Params} later in Section~\ref{sec:result}), we have found the optimal strategies $s^*$ and $f^*$ for the providers, as a function of government's strategy $\xi$. And these strategies appear to be continuous functions of $\xi$.  Hence the government's objective function from Section~\ref{sec:GovtProb} would be a continuous function of $\xi$.  Since that problem has a compact domain, a basic theorem from analysis~\cite[Theorem~7, p.~142]{Analysis_1} guarantees the existence of an optimum $\xi^*$.  This $\xi^*$ would then be the government's part of the perfect Nash equilibrium.

\section{Numerical Results}
\label{sec:result}
In the previous section, a simplified model with two providers and two regions was considered where the two providers' objective functions \eqref{eq:provider1} and \eqref{eq:provider2} were maximized by optimizing the subsidy amounts $s_{11}, s_{12}, s_{21}$ and $s_{22}$, and the fees $f_1$ and $f_2$. In this section, we first show the convergence of our simulations for the case of  two regions and two providers. Then, we analyze the characteristics of this model using the numerical equilibrium solutions. While we develop some insights into the subsidization model, we also validate the analytical equilibrium solutions derived in the previous section.

\subsection{Simulation Convergence to Equilibrium}
Due to complexity of the providers' problems represented in \eqref{eq:provider1} and \eqref{eq:provider2}, an analytical proof for the existence of a Nash equilibrium was deemed infeasible. However, our extensive simulations of the game converged and were based on \emph{best response algorithm}. Therefore, we guarantee that the simulations converge to an equilibrium. A two player game based on the best response algorithm is designed in Matlab for the case of two providers and two regions. During each iteration of the algorithm,
\begin{enumerate}
\item Provider~1's parameters $s_{11}, s_{12} \mbox{ and } f_1$ are optimized by solving the simultaneous equations \eqref{eq:provider1-foc1}--\eqref{eq:provider1-foc3};
\item Provider~2's parameters $s_{21}, s_{22} \mbox{ and } f_2$ are optimized by solving the simultaneous equations \eqref{eq:provider2-foc1}--\eqref{eq:provider2-foc3};
\item The optimized parameters evaluated in the previous two steps are used in \eqref{eq:provider1} and \eqref{eq:provider2} to compute the providers' objective values.
\end{enumerate}
To check for convergence, the evaluated objective values in the current iteration are compared to their respective values in the previous iteration. If the differences are smaller than a threshold $\epsilon$, convergence is attained.
\begin{table}[htp]
\caption{Input Parameter Settings}
\label{tab:InputParams}
\centering
\begin{tabular}{|l|l|}
\hline
\emph{Parameter}  & \emph{Value} \\
\hline \hline
$\xi_1, \xi_2$    & $262, 738$ \\ \hline
$\beta$ 		  & $76$\\ \hline
$n_1, n_2$ 		  & $26, 744$ \\ \hline
$\gamma,\epsilon$ & $0.05, 10^{-3}$ \\ \hline
\end{tabular}
\end{table}
\begin{figure}[htp]
\vspace{-3mm}
\begin{center}
\includegraphics[width = 3.2 in]{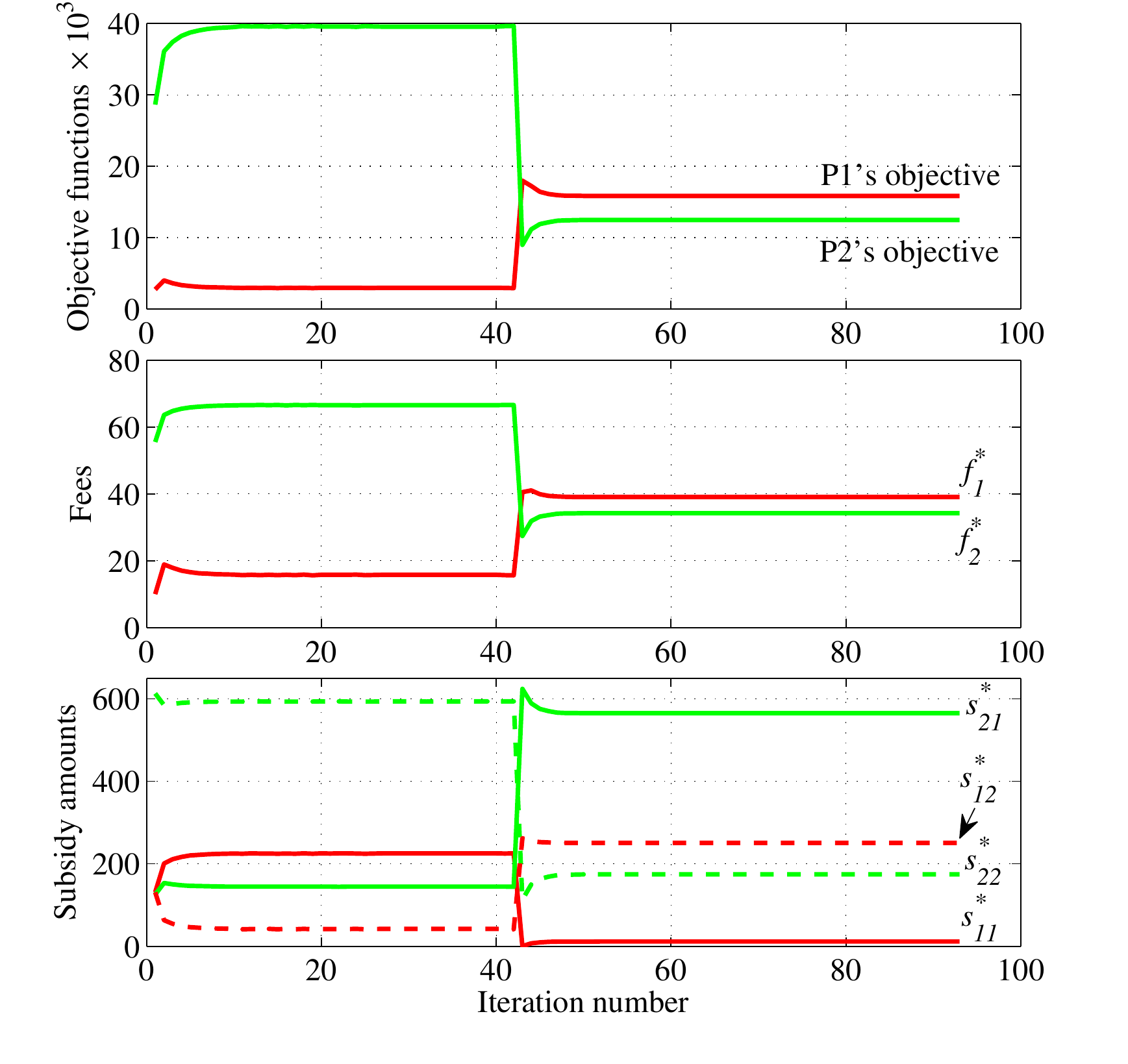}
\end{center}
\vspace{-5mm}
\caption{The convergence to equilibrium using best response algorithm. At beginning of the simulation, the system parameters were all set to zeros: $f_1=f_2=s_{11}=s_{12}=s_{21}=s_{22}=0$.}
\label{fig:Iterations}
\end{figure}

The best response simulator described above was run with arbitrary input parameter settings as shown in Table~\ref{tab:InputParams}. With these settings, the simulator took 93 iterations to converge. Behavior of the system parameters and the objective functions during the iterations are shown in Fig.~\ref{fig:Iterations}.

The number of iterations required for convergence can vary depending on the input parameter settings. To obtain the statistics of the number of iterations, the simulator was run 10,000 times with random input parameters. For each run, the input parameters $\beta, n_1$ and $n_2$ were generated using uniformly distributed random integers, within the range of values specified in Table~\ref{tab:ParamsRange} for each parameter. The parameters $\xi_1$ and $\xi_2$ were generated such that $\xi_1 + \xi_2 = 1000 = \xi$, the total budget of the government. While $\xi_1$ samples were generated using uniformly distributed random integers, the corresponding $\xi_2$ samples were calculated using $\xi_2 = \xi-\xi_1$.
\begin{table}[htp]
\caption{Range of Input Parameters}
\label{tab:ParamsRange}
\centering
\begin{tabular}{|l|l|}
\hline
\emph{Parameter} & \emph{Range} \\
\hline \hline
$\xi_1, \xi_2$ & $[50, 950]$ \\ \hline
$\beta$ 		& $[30, 200]$\\ \hline
$n_1, n_2$ 		& $[20, 1000]$ \\ \hline
\end{tabular}
\end{table}

The resulting distribution of the number of iterations is described in Table~\ref{tab:NoIteStatistics}. Most of the simulation runs (91.41\% of 10,000) converged within 15 iterations, while some of the simulation runs (3.88\%) took 16 to 99 iterations to converge. The remaining 4.71\% of the simulation runs took 100 or more iterations to converge.
\begin{table}[!t]
\caption{}
\label{tab:NoIteStatistics}
\centering
\begin{tabular}{|l|l|}
\hline
\emph{Iterations} & \emph{Number of occurrences}\\ \hline \hline
$\leq$ 15 & 9141\\ \hline
16-99 & 388\\ \hline
$\geq$ 100 & 471\\ \hline
\end{tabular}
\end{table}

\subsection{Numerical Results for Optimum Subsidy}
Assume that the government has a total subsidy budget of $\xi = 1000$, which has to be shared among the two providers such that $\xi_1 + \xi_2 = \xi = 1000$. Also assume that the other fixed parameters are set according to Table~\ref{tab:FixedParams}. As shown previously, equilibrium values ($s^*_{jk}$ and $f^*_j$) for the simplified subsidy model can be determined through the convergence of best response simulations as illustrated in Fig.~\ref{fig:Iterations}. The equilibrium values can also be obtained by numerically solving the eight simultaneous equations \eqref{eq:provider1-foc1}--\eqref{eq:provider1-foc3} and \eqref{eq:provider2-foc1}--\eqref{eq:provider2-foc3}. However, approximate equilibrium values can be easily obtained using the closed form expressions \eqref{s_11_approx}--\eqref{eq:f2} derived in the previous section, which also highlight the explicit dependence of optimum subsidy and fee values on the other key parameters.

\begin{table}[htp]
\caption{Fixed Parameters}
\label{tab:FixedParams}
\centering
\begin{tabular}{|l|l|}
\hline
\emph{Parameter} & \emph{Value} \\
\hline \hline
$\beta$ 		& $30$\\ \hline
$n_1, n_2$ 		& $40, 80$ \\ \hline
$\gamma$ 		& $0.05$ \\ \hline
\end{tabular}
\end{table}
\begin{figure}[htp]
\vspace{-3mm}
\begin{center}
\includegraphics[width = 3.2 in, trim = 0 10mm 0 0]{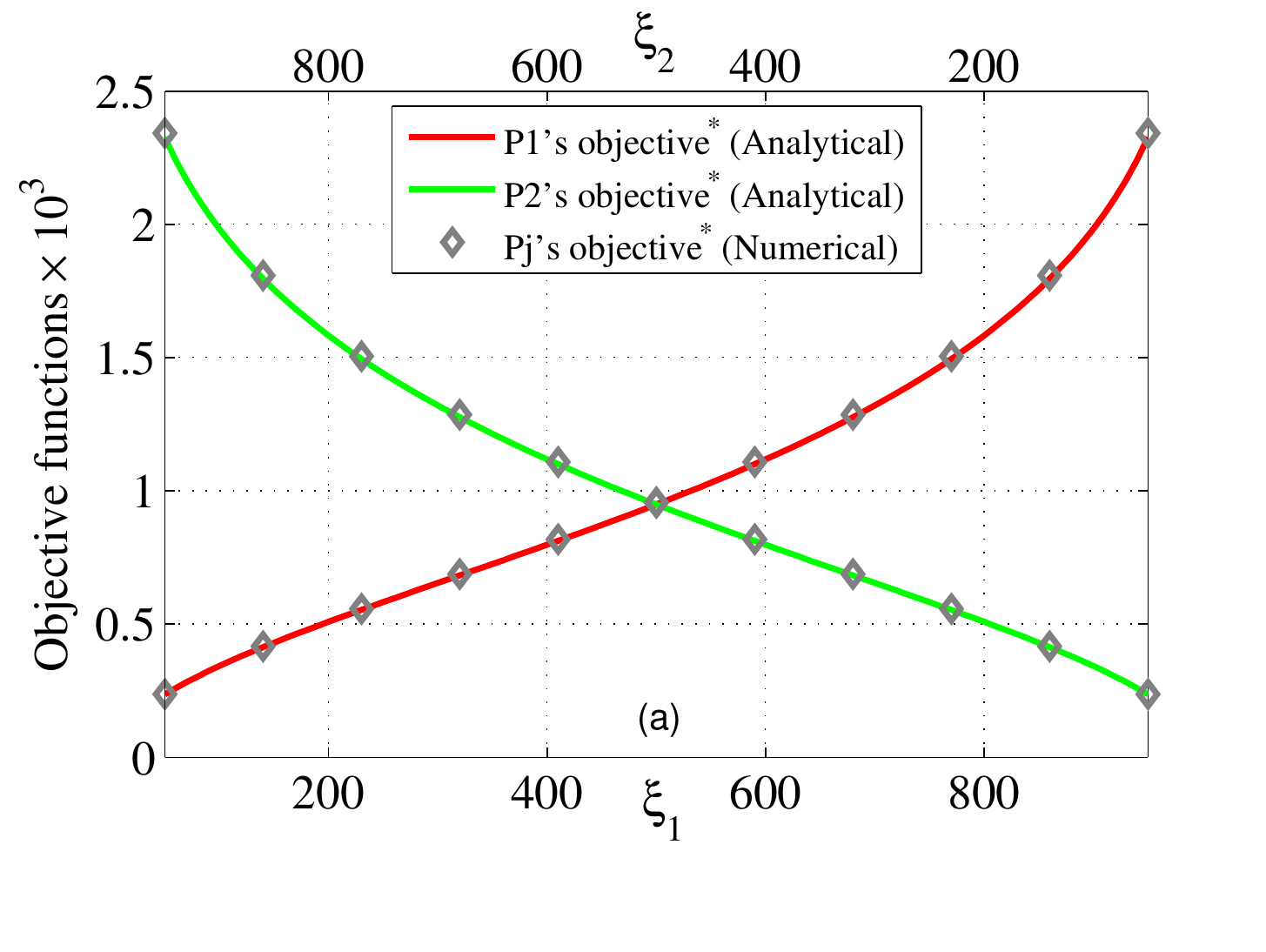}
\includegraphics[width = 3.2 in, trim = 0 10mm 0 0]{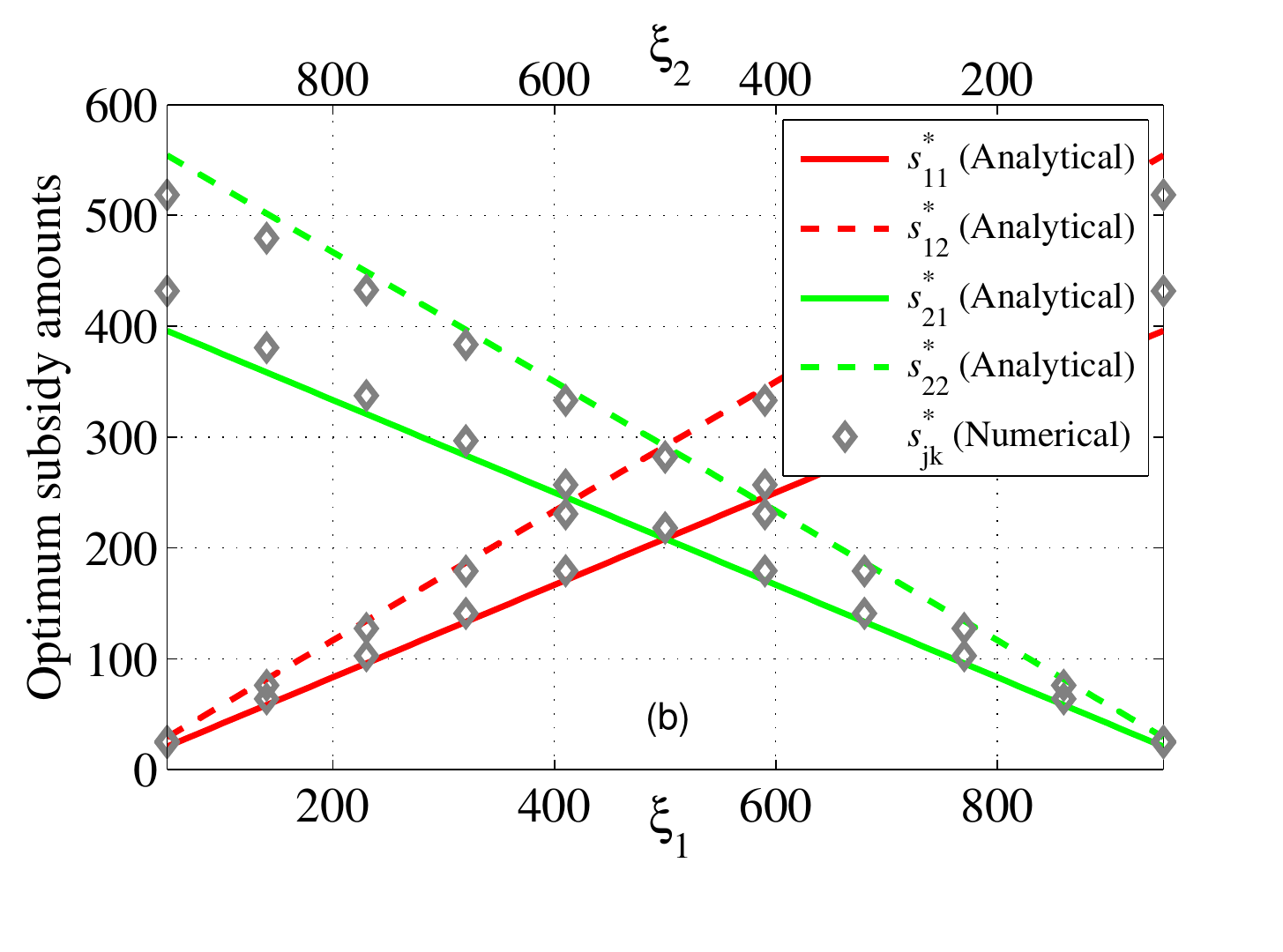}
\end{center}
\vspace{-5mm}
\caption{(a) Objective functions of the providers with optimum parameters; (b) Optimum subsidy amounts spent by the providers. Objective function of a provider increases if the government allocates more subsidy amount to that provider. Optimum subsidy amounts are proportional to the subsidy amounts allocated by government to the providers.}
\label{fig:Obj_sjk_vs_xi1_xi2}
\end{figure}
In here, we plot analytical results using the expressions \eqref{s_11_approx}-\eqref{s_22_approx} for the optimum subsidy amounts ($s^*_{jk}$). To validate the accuracy of these analytical expressions, numerical solution of the eight simultaneous equations is also plotted in the same graph. Under this scenario, analytical and numerical \emph{optimum values} of the providers' objective functions are plotted with respect to the government subsidies $\xi_1$ and $\xi_2$ in Fig.~\ref{fig:Obj_sjk_vs_xi1_xi2}(a). It can be seen that the objective function of a provider increases if the government allocates higher subsidy to that provider. Consequently, the objective function of the other provider decreases. Therefore, the government can easily avoid monopolization in the market by controlling the allocations of the subsidy amount. For example, government can allocate higher subsidy amount to small providers so that they get a fair chance to compete with large providers. Note that the analytical and numerical results in Fig.~\ref{fig:Obj_sjk_vs_xi1_xi2} match very closely.

\begin{figure}[htp]
\begin{center}
\includegraphics[width = 3.2 in]{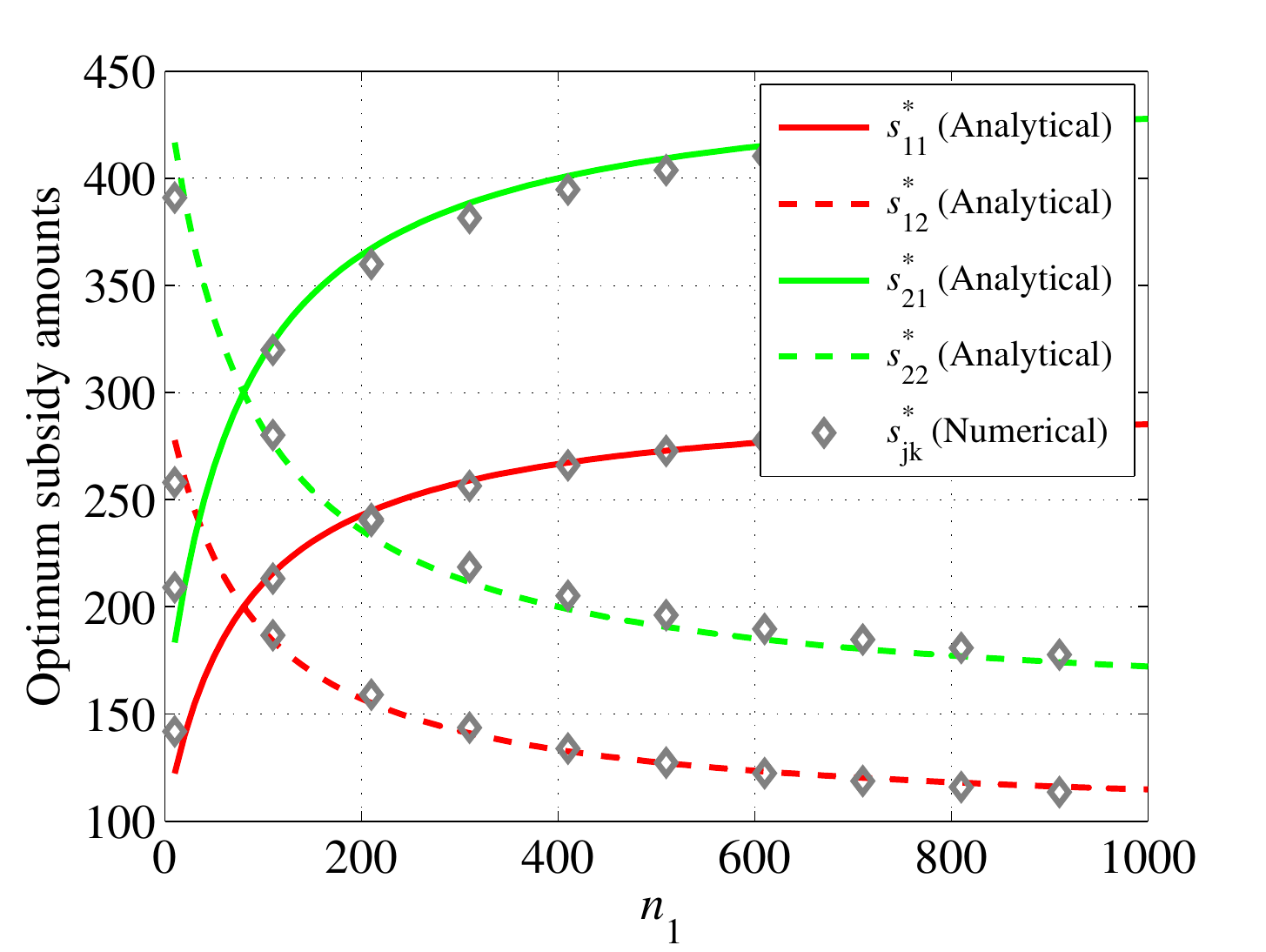}
\end{center}
\vspace{-5mm}
\caption{Optimum subsidy amounts spent by the two providers versus the number of users in region 1, and fixed number of users in region 2. It can be noted that providers spend higher amounts in the region with more number of users.}
\label{fig:Sxx_vs_n1}
\end{figure}

Fig.~\ref{fig:Obj_sjk_vs_xi1_xi2}(b) shows the optimum subsidy amounts to be spent by the providers in each region. The optimum subsidy amounts spent are proportional to the subsidy amounts allocated by the government to the providers. The providers should always spend higher amounts in the regions with higher user densities to offer good quality of service and at the same time be more profitable. Accordingly, in Fig.~\ref{fig:Obj_sjk_vs_xi1_xi2}(b) it can be observed that since $n_2 > n_1$, the two providers spend higher amounts ($s_{12}$ and $s_{22}$) in region~2 when compared to region~1. This is also illustrated in Fig.~\ref{fig:Sxx_vs_n1}, where $n_1$ is varied from $10$ to $1000$ while $n_2$ is fixed to 80, and the government subsidies $\xi_1$ and $\xi_2$ are fixed to $400$ and $600$, respectively. As $n_1$ increases, both the providers spend higher amounts in region~1 and lower amounts in region 2. This shows that our subsidy framework will result in a balanced investment across regions and thus avoid over-investment into a subset of the regions.

Another point to note is, according to \eqref{s_11_approx}--\eqref{s_22_approx}, $s_{11}^*+s_{12}^*=\xi_1$ and $s_{21}^*+s_{22}^*=\xi_2$. It means that a provider can maximize its revenue only if it spends the entire subsidy amount received from the government in the regions appropriately. The provider cannot get maximum benefit if it spends only partial amount of the received subsidy.

\subsection{Numerical Results for Optimum Fee}
\begin{figure}[htp]
\begin{center}
\includegraphics[width = 3.2 in, trim = 0 12mm 0 0]{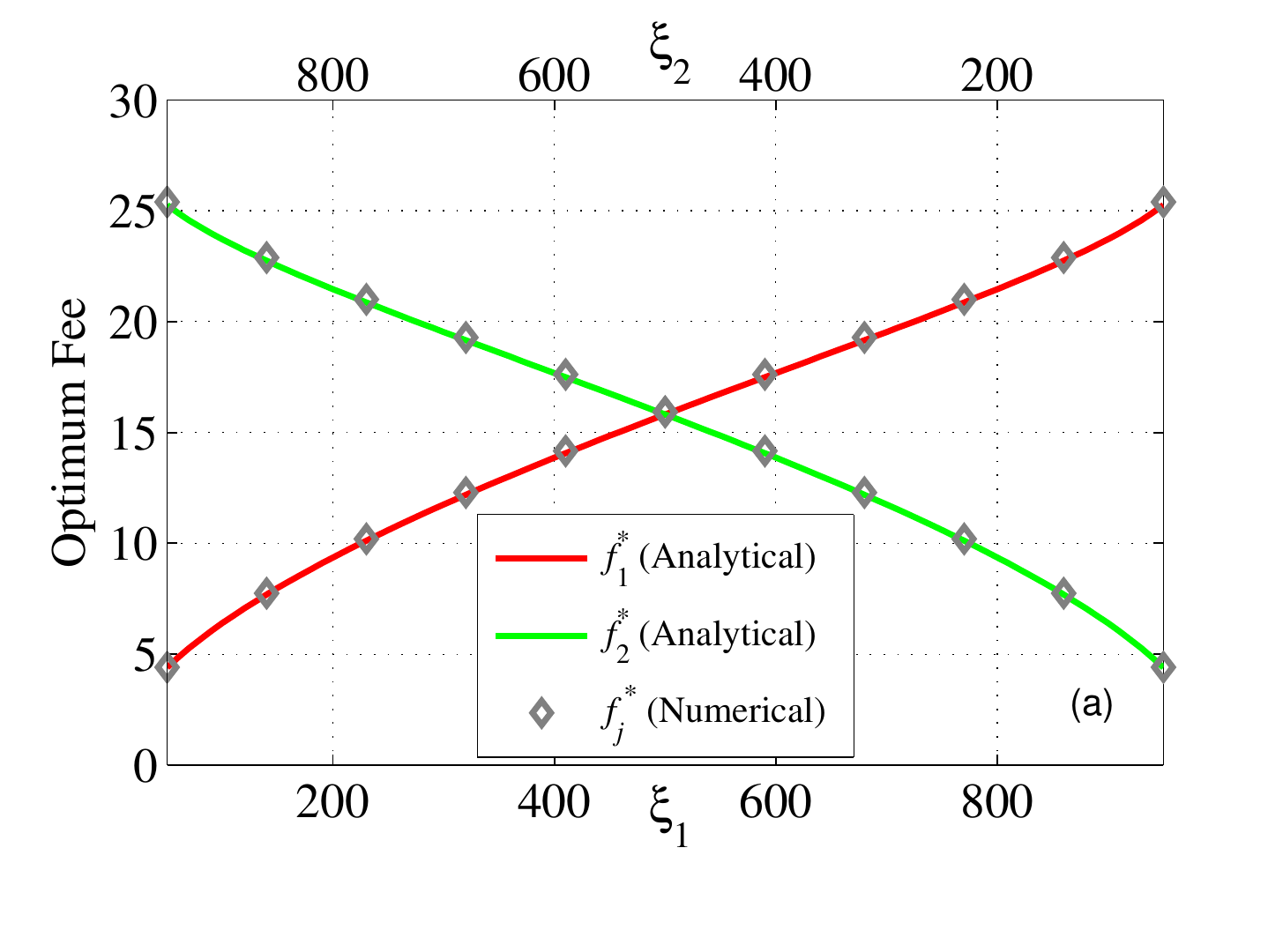}
\includegraphics[width = 3.2 in]{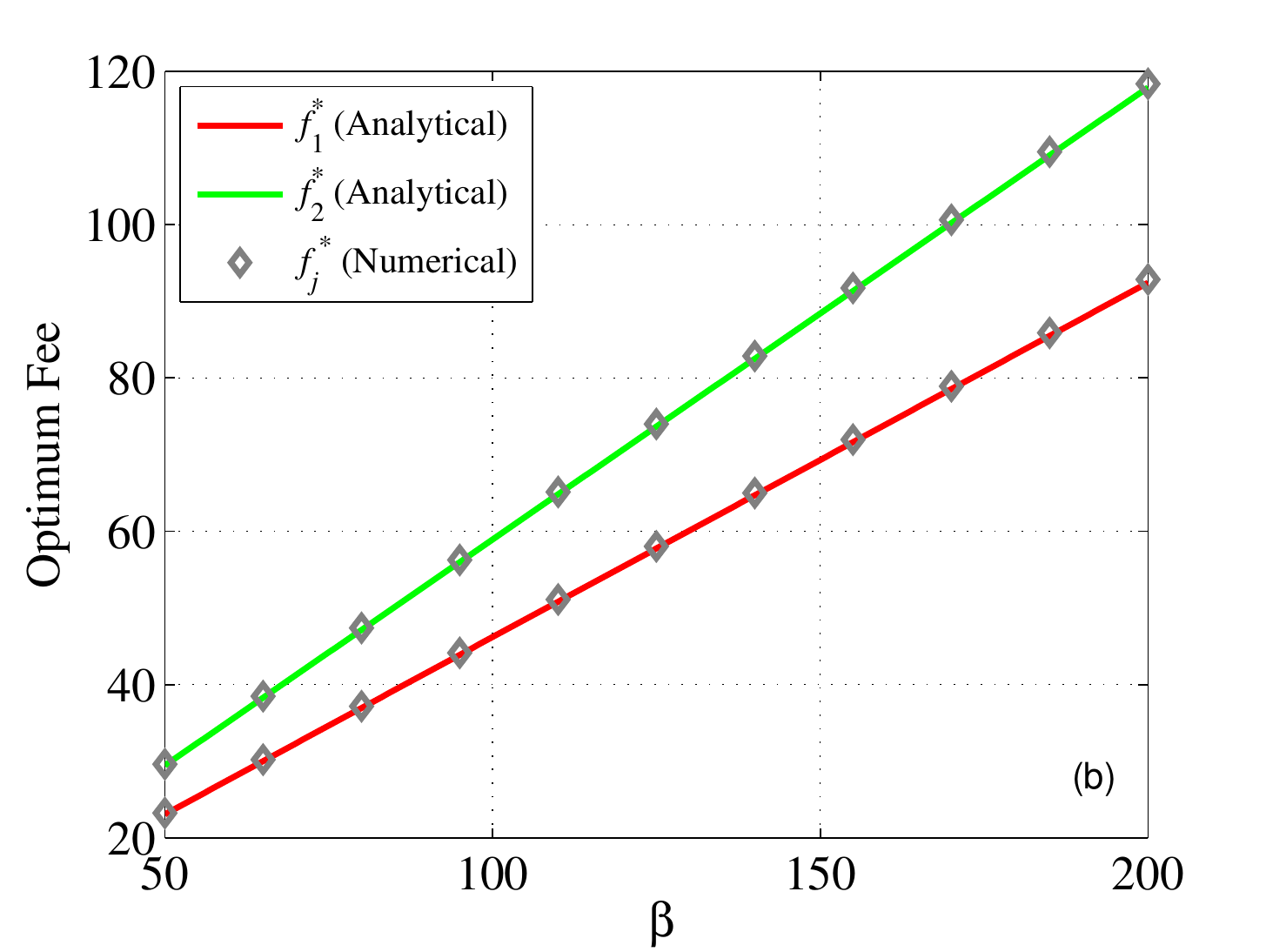}
\end{center}
\vspace{-5mm}
\caption{Optimum fee charged to users versus (a) the allocated subsidy amount; (b) the number of calls made by a customer. Optimum fee significantly depends on the subsidy allocated to the provider, and is proportional to $\beta$.}
\label{fig:Fx_vs_Params}
\end{figure}
A comparison between the analytical and numerical results for the optimum fee are shown in Fig.~\ref{fig:Fx_vs_Params}. The optimum fee that a provider should charge to its customers significantly depends on the total subsidy amount $\xi_j$ it receives from the government, as shown in Fig.~\ref{fig:Fx_vs_Params}(a). If government increases the subsidy amount to a provider, it enables that provider to improve its quality of service, and hence increase the fee, and eventually to improve its revenue. Likewise, if the government provides higher subsidy to small operators, they will be able to compete with large providers. As shown in Fig.~\ref{fig:Fx_vs_Params}(b), the fee charged to a user is proportional to the number of calls made by the customer. Here, the fee charged by provider~2 is more than that of provider~1 because of higher subsidy amount allocated to provider~2, that is, $\xi_1 = 400$ and $\xi_2 = 600$.

\begin{figure}[htp]
\begin{center}
\includegraphics[width = 3.2 in]{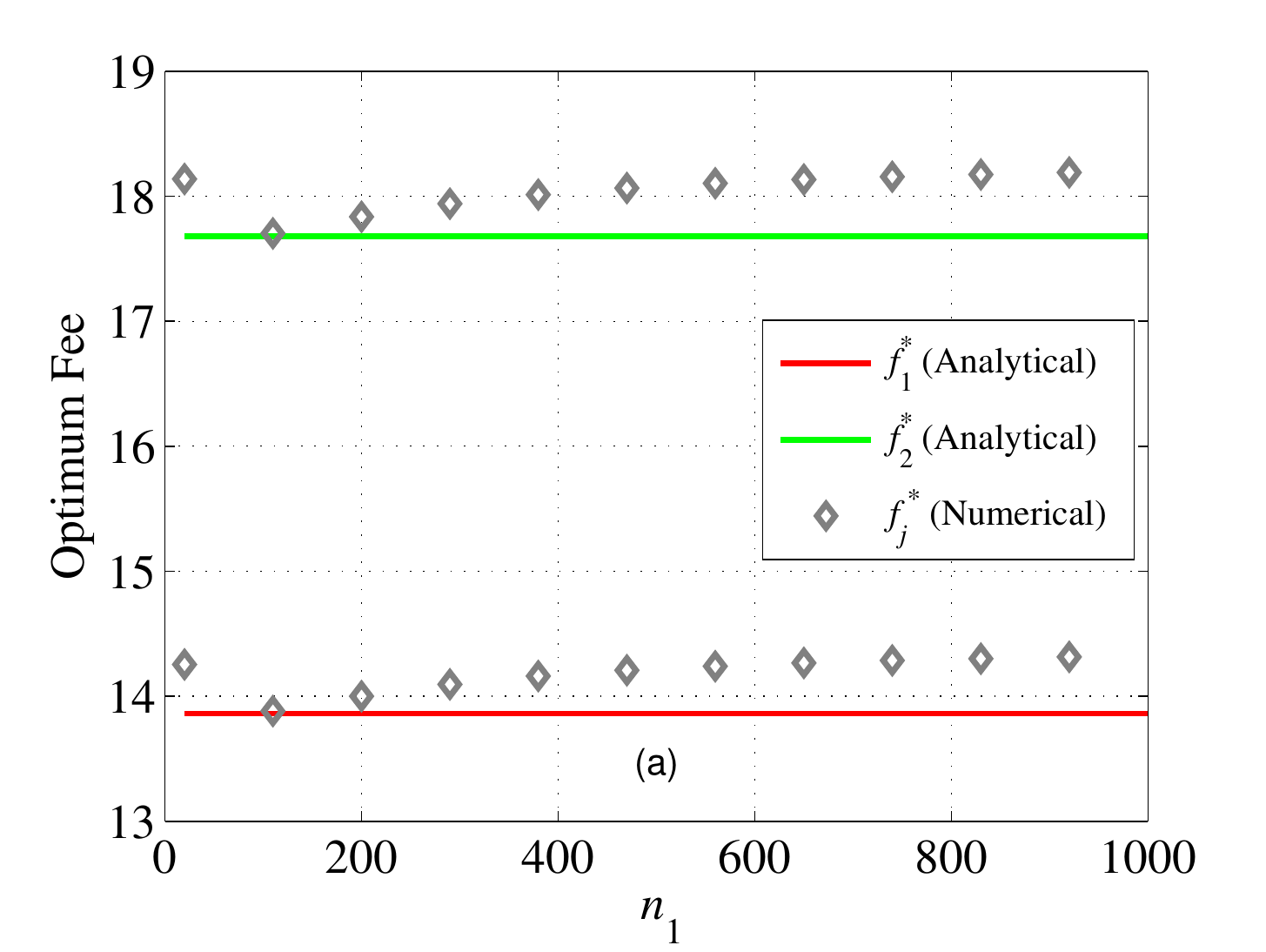}
\includegraphics[width = 3.2 in]{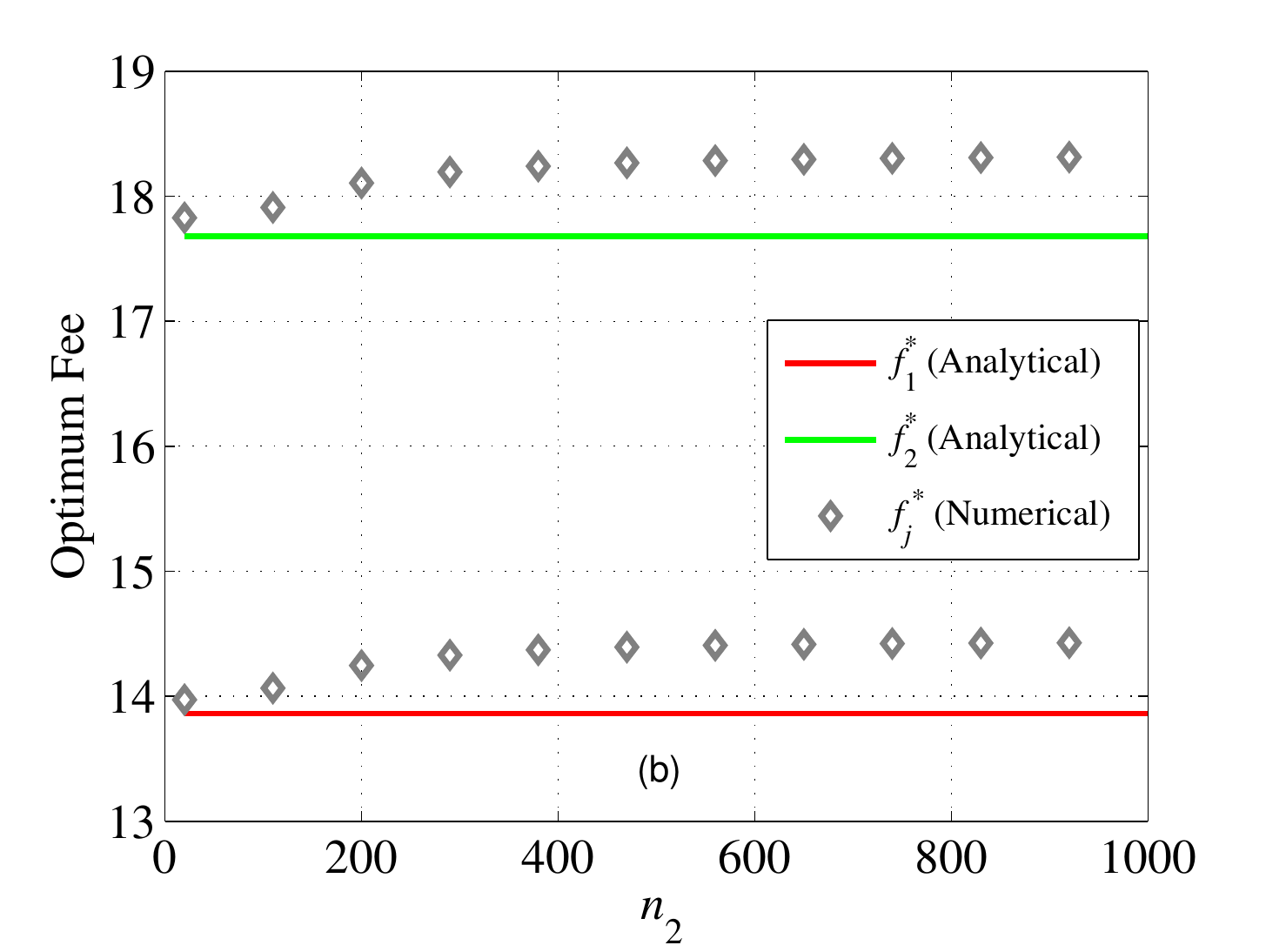}
\end{center}
\vspace{-5mm}
\caption{Optimum fee charged to users versus the number of users (a) in region~1; (b) in region~2. Variations of optimum fees is negligible with respect to $n_1$ and $n_2$.}
\label{fig:Fx_vs_n1_n2}
\end{figure}
As mentioned in Section~\ref{subsec:Fj}, the number of users in each region $n_1$ and $n_2$ has very little effect on the fees charged by the providers in each region. Accordingly, the numerical plots of $f_1$ and $f_2$ in Fig.~\ref{fig:Fx_vs_n1_n2} are almost independent of $n_1$ and $n_2$. The analytical plots which are based on the assumption $n_1=n_2=n$ closely match with the numerical plots.

%
%

\section{Conclusion}
\label{sec:summary}
In this paper, we have proposed a spectrum market where sharing is promoted explicitly by the government. The government offers subsidy support to the wireless operators and requires a performance metric to be reported. We considered this metric to be ``the number of outside customers'' served by the operator, where ``outside customer'' means a customer who is not subscribed to that particular operator. This way, the operators are motivated to invest their subsidy support into regions where likelihood of reaching and serving outside customers is higher. This, in turn, increases the aggregate coverage area and the overall signal quality the customers receive. Both of these end results are appealing to the government and are the main reasons why the government should provide subsidy.

We have proposed a game-theoretic framework to model the interactions between the government and the providers. We have derived first-order conditions for a perfect Nash equilibrium. We have further specialized the general problem for a two-provider and two-region scenario. For this simplified case, we have shown existence of such an equilibrium based on the best response algorithm, and also derived closed form approximations to the providers' equilibrium conditions.

Our results have shown how the government should influence the spectrum sharing by controlling the subsidy amount allocated to each provider. We have also studied how the providers can distribute the received subsidy amount into different regions to serve maximum number of outside customers and eventually maximize their revenues.

We have highlighted various benefits of a spectrum market with performance-based subsidy. Such a spectrum market will be attractive to the operators and discourage monopolization. Further more, small providers with relatively smaller infrastructure can survive in this type of market and successfully compete against large operators.

\appendix[Proof of Theorem 1]
\label{App:Fee_opt_expr}
In this appendix, closed form expressions for the optimum fee $f^*_1$ and $f^*_2$ are derived. By applying the assumption $n_1=n_2=n$ in \eqref{s_11_approx}--\eqref{s_22_approx}, we get
\begin{align}
s^*_{11}=s^*_{12}=\xi_1/2 \equiv s^*_1,\\
s^*_{21}=s^*_{22}=\xi_2/2 \equiv s^*_2.
\end{align}
By substituting these two results in \eqref{eq:provider1} and \eqref{eq:provider2}, they can be simplified respectively as
\begin{align}
\smash{\max_{f_1}}\ \ & \frac{2 f_1 n}{1 + \frac{\gamma \beta \sqrt{s^*_{2}} - f_2}{\gamma \beta \sqrt{s^*_{1}} - f_1}} + {\xi\sqrt{s^*_{1}}  \over \sqrt{s^*_{1}} + \sqrt{s^*_{2}} } - 2 s^*_{1} \label{eq:provider1_smpl} \\
\rm{s.~t.,} \ \ & f_1 \ge 0, \label{eq:provider1-cons2_smpl}
\end{align}
and
\begin{align}
\smash{\max_{f_2}}\ \ & \frac{2 f_2 n}{1 + \frac{\gamma \beta \sqrt{s^*_{1}} - f_1}{\gamma \beta \sqrt{s^*_{2}} - f_2}} + {\xi\sqrt{s^*_{2}}  \over \sqrt{s^*_{1}} + \sqrt{s^*_{2}} } -2 s^*_{2} \label{eq:provider2_smpl} \\
\rm{s.~t.,}\ \ & f_2 \ge 0. \label{eq:provider2-cons2_smpl}
\end{align}
Then, the first order conditions with respect to $f_1$ and $f_2$ can be respectively derived to be:
\begin{align}
(\gamma\beta\sqrt{s^*_1}-f_1)^2 + (\gamma\beta\sqrt{s^*_2}-f_2) (\gamma\beta\sqrt{s^*_1}-2 f_1) &= 0, \label{eq:f1_cond}\\
(\gamma\beta\sqrt{s^*_2}-f_2)^2 + (\gamma\beta\sqrt{s^*_1}-f_1) (\gamma\beta\sqrt{s^*_2}-2 f_2) &= 0. \label{eq:f2_cond}
\end{align}
For simplicity, substitute $x=\gamma\beta\sqrt{s^*_1}-f_1$ and $y=\gamma\beta\sqrt{s^*_2}-f_2$ in \eqref{eq:f1_cond} and \eqref{eq:f2_cond} to get:
\begin{align}
x^2 + y (2x-\gamma\beta\sqrt{s^*_1}) &= 0, \label{eq:f1_cond_2}\\
y^2 + x (2y-\gamma\beta\sqrt{s^*_2}) &= 0. \label{eq:f2_cond_2}
\end{align}
Since $\beta$, $s^*_1$ and $s^*_2$ are known, if we solve for $x$ and $y$, then $f_1$ and $f_2$ can be solved easily. Writing \eqref{eq:f1_cond_2} and \eqref{eq:f2_cond_2} in terms of $y$ and $x$ respectively, we get:
\begin{align}
y &= \frac{-x^2}{2x-\gamma\beta\sqrt{s^*_1}},\label{eq:y_sol}\\
x &= \frac{-y^2}{2y-\gamma\beta\sqrt{s^*_2}}.\label{eq:x_sol}
\end{align}
By substituting \eqref{eq:y_sol} and \eqref{eq:x_sol} in \eqref{eq:f2_cond_2} and \eqref{eq:f1_cond_2} respectively, and simplifying, we get:
\begin{align}
x^3 - \frac{2\gamma\beta\left(\sqrt{s^*_1} - 2\sqrt{s^*_2}\right)}{3}x^2 - \frac{4\gamma^2\beta^2\sqrt{s^*_1 s^*_2}}{3}x &\nonumber\\
+ \frac{\gamma^3\beta^3 s^*_1 \sqrt{s^*_2}}{3} &= 0,\label{eq:x_poly}\\
y^3 - \frac{2\gamma\beta\left(\sqrt{s^*_2} - 2\sqrt{s^*_1}\right)}{3}y^2 - \frac{4\gamma^2\beta^2\sqrt{s^*_1 s^*_2}}{3}y &\nonumber\\
+ \frac{\gamma^3\beta^3 s^*_2 \sqrt{s^*_1}}{3} &= 0.\label{eq:y_poly}
\end{align}
It can be observed that \eqref{eq:x_poly} and \eqref{eq:y_poly} are 3rd order polynomials represented in the standard form. Consider the polynomial in \eqref{eq:x_poly}. By substituting $x = t + 2\gamma\beta(\sqrt{s^*_1} - 2\sqrt{s^*_2})/9$, the quadratic term can be eliminated and the polynomial can be expressed in a more compact form as:
\begin{align}
t^3 - A t - B = 0, \label{eq:DepressedPoly}
\end{align}
\begin{align}
\mbox{where, }A &= \frac{4\gamma^2\beta^2\left[9\sqrt{s^*_1 s^*_2}+\left(\sqrt{s^*_1}-2\sqrt{s^*_2}\right)^2\right]}{27},\\
B &= \frac{\gamma^3\beta^3}{27^2}\left(\splitfrac{16 s^*_1 \sqrt{s^*_1} - 240 s^*_2 \sqrt{s^*_1}} {- 123 s^*_1 \sqrt{s^*_2} - 128 s^*_2\sqrt{s^*_2}}\right).
\end{align}
Using Viete's method, the three real roots of \eqref{eq:DepressedPoly} can be found as follows:
\begin{align}
\hspace{-1mm}t^*_k = 2\sqrt{\frac{A}{3}}\cos\left(\frac{1}{3}\cos^{-1}\sqrt{\frac{27B^2}{4A^3}}+\frac{(3-2k)\pi}{3}\right),
\end{align}
for k = 0,1,2. Then, the roots of $x$ in \eqref{eq:x_poly} can be expressed as:
\begin{align}
x^*_k =& t^*_k + \frac{2\gamma\beta(\sqrt{s^*_1} - 2\sqrt{s^*_2})}{9},\nonumber\\
=& 2\sqrt{\frac{A}{3}}\cos\left(\frac{1}{3}\cos^{-1}\sqrt{\frac{27B^2}{4A^3}}+\frac{(3-2k)\pi}{3}\right)\nonumber\\
&+ \frac{2\gamma\beta(\sqrt{s^*_1} - 2\sqrt{s^*_2})}{9}.
\end{align}
Finally, the roots of $f_1$ in \eqref{eq:f1_cond} can be expressed as
\begin{align}
f^*_{1k} =& \gamma\beta\sqrt{s^*_1} - x^*_k,\\
=& \frac{\gamma\beta(7\sqrt{s^*_1}+4\sqrt{s^*_2})}{9}\nonumber\\
&-2\sqrt{\frac{A}{3}}\cos\left(\frac{1}{3}\cos^{-1}\sqrt{\frac{27B^2}{4A^3}}+\frac{(3-2k)\pi}{3}\right).
\end{align}
Out of the three solutions $f^*_{1k},$ for $k=0,1, \mbox{ and } 2,$ one solution should be chosen which matches with the numerical results.
\begin{figure}[htp]
\begin{center}
\includegraphics[width = 3.2 in]{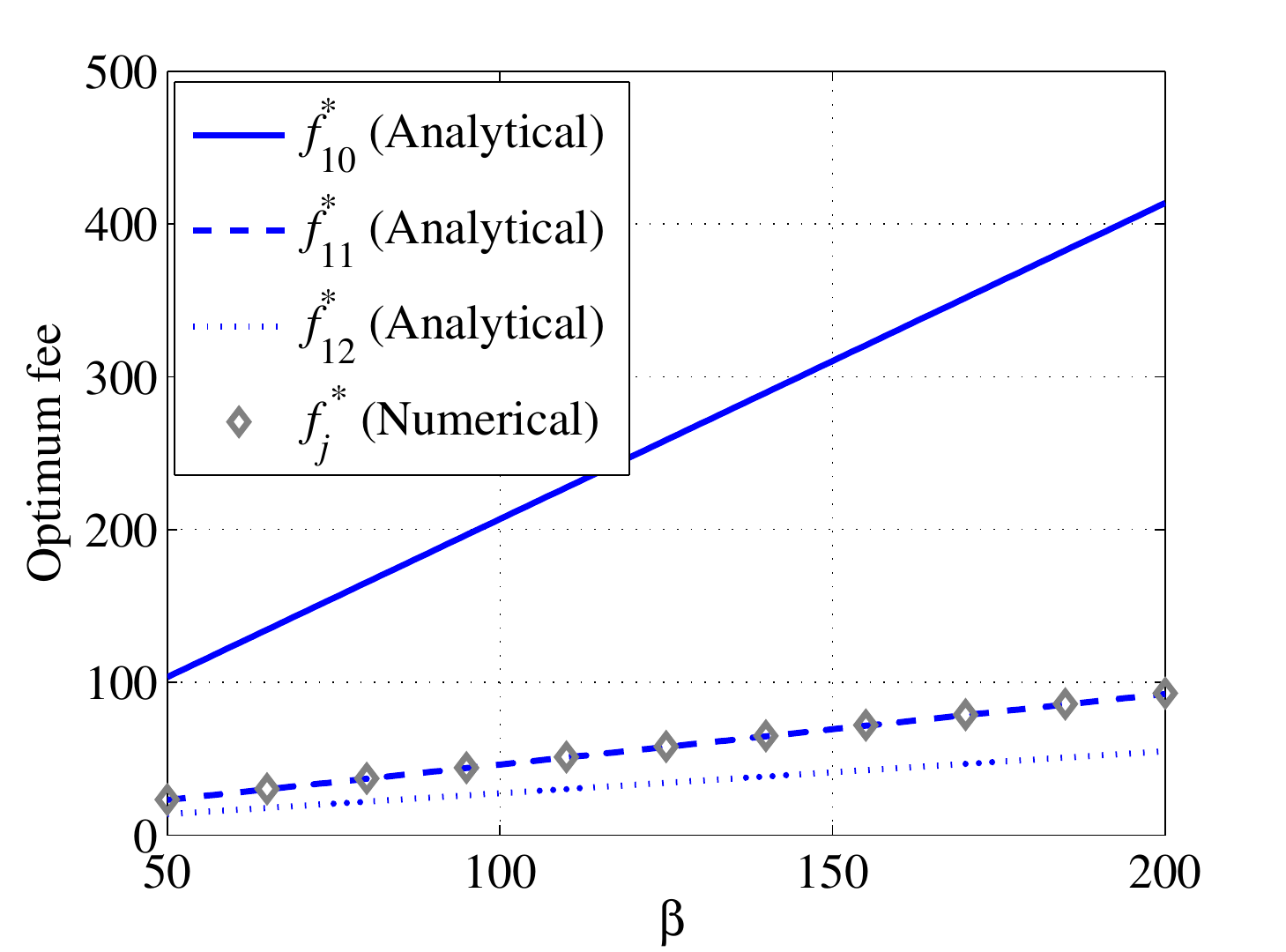}
\end{center}
\vspace{-5mm}
\caption{Solutions of $f_1$ and comparison with numerical solution.}
\label{fig:F1_sols}
\end{figure}
A comparison of the three solutions with the numerical result is shown in Fig.~\ref{fig:F1_sols}. The solution with $k=1$ matches with the numerical solution. Therefore, the optimum fee for provider 1 can be written as
\begin{align}
f^*_1 =& \frac{\gamma\beta(7\sqrt{s^*_1}+4\sqrt{s^*_2})}{9}\nonumber\\
&-2\sqrt{\frac{A}{3}}\cos\left(\frac{1}{3}\cos^{-1}\sqrt{\frac{27B^2}{4A^3}}+\frac{\pi}{3}\right).
\end{align}
Using a similar approach as described in this appendix, a closed form expression of the solution $f^*_2$ can also be derived as follows:
\begin{align}
f^*_2 =& \frac{\gamma\beta(7\sqrt{s^*_2}+4\sqrt{s^*_1})}{9}\nonumber\\
&-2\sqrt{\frac{C}{3}}\cos\left(\frac{1}{3}\cos^{-1}\sqrt{\frac{27D^2}{4C^3}}+\frac{\pi}{3}\right),
\end{align}
where
\begin{align}
C &= \frac{4\gamma^2\beta^2\left[9\sqrt{s^*_2 s^*_1}+\left(\sqrt{s^*_2}-2\sqrt{s^*_1}\right)^2\right]}{27},\\
D &= \frac{\gamma^3\beta^3}{27^2}\left(\splitfrac{16 s^*_2 \sqrt{s^*_2} - 240 s^*_1 \sqrt{s^*_2}} {- 123 s^*_2 \sqrt{s^*_1} - 128 s^*_1\sqrt{s^*_1}}\right).
\end{align}

\section*{Acknowledgment}
This research was supported in part by the U.S. National Science Foundation under the grants AST-1443999, AST-1443913, AST-1444077 and CNS-1321069.

\bibliographystyle{IEEEtran}
\bibliography{./bib/networking,./bib/new}

%
%
%
%
%
%
%

\end{document}